\ifdefined\pdfpagewidth\else\newdimen\pdfpagewidth\fi
\ifdefined\pdfpageheight\else\newdimen\pdfpageheight\fi
\ifdefined\pdfoutput\else\newcount\pdfoutput\pdfoutput=1\fi
\documentclass[paper]{JFM-FLM_Au-polished}

\usepackage{graphicx}
\usepackage{subcaption}
\usepackage{epstopdf,epsfig}
\IfFileExists{newtxtext.sty}{\usepackage{newtxtext,newtxmath}}{}
\usepackage{natbib}
\usepackage{hyperref}
\usepackage{amsmath,bm,mathtools}
\IfFileExists{siunitx.sty}{\usepackage{siunitx}}{%
  \newcommand{\SI}[2]{##1\,\mathrm{##2}}%
}
\usepackage{booktabs,tabularx,array}
\usepackage{xcolor}
\usepackage[nameinlink,noabbrev]{cleveref}
\usepackage{enumitem}
\usepackage{placeins}

\makeatletter
\def\ps@reviewplain{%
  \let\@mkboth\@gobbletwo
  \def\@oddhead{\hfil{\itshape\@righttitle}\hfil}%
  \def\@evenhead{\hfil{\itshape\@lefttitle}\hfil}%
  \def\@oddfoot{\hbox to \textwidth{\hfill{\cppagefont\thepage}}}%
  \def\@evenfoot{\hbox to \textwidth{{\cppagefont\thepage}\hfill}}%
  \def\sectionmark##1{}\def\subsectionmark##1{}%
}
\makeatother

\hypersetup{colorlinks=true,urlcolor=blue,citecolor=black,linkcolor=black}
\graphicspath{{figures/}}
\newcommand{\KnD}{Kn_D}
\newcommand{\tstar}{t^*}
\newcommand{\scenter}{s_{50}}
\newcommand{\deltaten}{\delta_{10\text{--}90}}
\newcommand{\mean}[1]{\overline{#1}}
\newcommand{\rms}[1]{#1_{\mathrm{rms}}}

\title{Noise-separated evidence for a slow collective displacement in a rarefied hypersonic bow-shock layer}
\lefttitle{Slow collective displacement of a rarefied bow shock}
\righttitle{A. Shoja-Sani and E. Roohi}
\corresau{Ehsan Roohi, \email{roohie@umass.edu}}

\author{Ahmad Shoja-Sani\aff{1}\and Ehsan Roohi\aff{2}}
\affiliation{
\aff{1}Department of Mechanical Engineering, Ferdowsi University of Mashhad, Mashhad 9177948974, Iran
\aff{2}Department of Mechanical and Industrial Engineering, University of Massachusetts Amherst, 160 Governors Drive, Amherst, MA 01003, USA
}

\begin{document}
\maketitle
\pagestyle{reviewplain}
\thispagestyle{reviewplain}

\begin{abstract}
Time-resolved direct simulation Monte Carlo (DSMC) fields are used to test whether a detached rarefied hypersonic bow shock contains a slow collective displacement that can be separated from correlated particle-sampling fluctuations. Mach-10 rotationally relaxing nitrogen flow over a circular cylinder is analysed for diameter-based Knudsen number $0.01\leq \KnD\leq1$, where $\KnD=\lambda_\infty/D$, $\lambda_\infty$ is the freestream mean free path and $D$ is the cylinder diameter. A density half-jump front is extracted on body-normal rays, unsupported solid-side points are excluded, and temporal coarse graining is performed before feature extraction. Persistent and sampling covariance components are compared using a penalized composite-fit score, design-scale cross-validation, block resampling, synthetic controls and complementary full-field matched filters. Corrected field proper orthogonal decomposition (POD) is high rank at every Knudsen number, yet a weak, same-signed angular displacement is resolved at $\KnD=0.01$ and $0.025$. Independent random-seed and simulator-particle-loading repeats recover the angular shape and relaxation time while the raw sampling variance changes with loading. Across the two resolved states the mean density layer broadens by $82\%$, while the angular shapes remain strongly aligned. Density and pressure recover the marker motion most strongly; the reduced Mach-number and translational-temperature participation at $\KnD=0.025$ is evidence consistent with moment-selective weakening, although observable-dependent signal-to-noise remains a possible contributor. The signal is interpreted as a low-pass bow-layer response embedded in broadband kinetic fluctuations, not as a newly discovered discrete oscillation or a demonstrated linear instability. The higher-Knudsen records are not sufficiently sensitive to establish physical disappearance.
\end{abstract}


\clearpage
\subsection{Nomenclature and conventions}

\begingroup
\small
\setlength{\tabcolsep}{5pt}
\renewcommand{\arraystretch}{1.08}

\noindent\textit{Physical and geometrical quantities.}\par\nobreak\vspace{2pt}
\noindent\begin{tabularx}{\textwidth}{@{}>{$}l<{$}X@{}}
\toprule
\text{Symbol} & \text{Definition} \\
\midrule
D,\ R=D/2 & Cylinder diameter and radius. \\
r,\ s=r-R & Distance from the cylinder centre and wall-normal distance from the cylinder surface. \\
\theta,\ \theta' & Polar angle and a second ray angle; the upstream stagnation line is at $\theta=180^\circ$. \\
t,\ \Delta t & Dimensional time and interval between output-block centres. \\
t^*=tU_\infty/D,\ \Delta t^*=\Delta t\,U_\infty/D & Dimensionless convective time and dimensionless output cadence. \\
\lambda_\infty,\ Kn_D=\lambda_\infty/D & Freestream mean free path and diameter-based Knudsen number. \\
U_\infty,\ M_\infty & Freestream speed and Mach number. \\
T,\ T_2,\ T_\infty,\ T_w & Temperature used in the equilibrium estimate, estimated post-shock temperature, freestream temperature and wall temperature. \\
\rho,\ p,\ M,\ T_{tr} & Density, pressure, local Mach number and translational temperature. \\
\rho_\infty,\ \rho_{\mathrm{rms}} & Freestream density and root-mean-square density fluctuation. \\
\rho_{up},\ \rho_{down} & Robust local upstream and downstream density plateaux on a body-normal ray. \\
\rho^*=(\rho-\rho_{up})/(\rho_{down}-\rho_{up}) & Normalized density transition used for front extraction. \\
s_{50},\ \delta_{10\text{--}90} & Density half-jump location and distance between the 10\% and 90\% density-transition levels. \\
\xi=(s-s_{50})/\delta_{10\text{--}90} & Shock-attached wall-normal coordinate. \\
q(s,\theta,t),\ \overline q,\ q'=q-\overline q & Generic macroscopic field, its temporal mean and its fluctuation. \\
q'_\perp & High-rank residual outside the retained displacement coordinate. \\
u_n,\ s_a,\ s_b & Local body-normal velocity and the two limits used to define a layer residence time. \\
\gamma,\ R_s & Specific-heat ratio and nitrogen-specific gas constant. \\
\theta_v,\ c_{v,v},\ x=\theta_v/T & Characteristic vibrational temperature, vibrational contribution to constant-volume specific heat and vibrational-temperature ratio. \\
\bottomrule
\end{tabularx}

\medskip\noindent\textit{Front displacement, covariance and modal quantities.}\par\nobreak\vspace{2pt}
\noindent\begin{tabularx}{\textwidth}{@{}>{$}l<{$}X@{}}
\toprule
\text{Symbol} & \text{Definition} \\
\midrule
a(\theta,t),\ a_m(t) & Local front-normal displacement and scalar angular-marker amplitude. \\
a_q(t) & Equivalent displacement amplitude obtained by projecting field $q$ onto its translation template. \\
g(\theta) & Normalized persistent angular displacement mode. \\
\Psi_q=-g\,\partial_s\overline q & Full-field translation template for observable $q$. \\
W,\ W_j,\ \langle u,v\rangle_W & Spatial weight field, its value at retained grid point $j$ and the associated discrete weighted inner product. \\
N_s,\ N_p & Number of stored snapshots and reference simulator-particle loading. \\
m,\ k & Number of consecutive fields in a coarse-graining group and integer lag; in $IC_c$, $k$ instead denotes the number of fitted scalar parameters. \\
A_m(\phi) & Variance-attenuation factor for averaging $m$ samples of a first-order autoregressive process. \\
\phi,\ \phi_p,\ \phi_n & Lag-one autoregressive coefficient and its persistent and sampling-component values. \\
C_m,\ C_p,\ C_n & Measured marker covariance after grouping $m$ fields, persistent covariance and correlated sampling covariance. \\
\lambda_1(C_p),\ \sigma_1=\sqrt{\lambda_1(C_p)} & Leading persistent-covariance eigenvalue and its modal standard deviation. \\
\tau_p^*=-\Delta t^*/\log\phi_p & Dimensionless persistent-memory time scale. \\
J,\ n_c & Composite fitting objective and number of retained summary terms. \\
IC_c,\ \Delta IC_c & Penalized composite-fit score and noise-only minus two-component score difference. \\
R_{cv},\ \epsilon_{proj} & Two-component/noise-only LOgSO-CV error ratio and relative positive-semidefinite projection correction. \\
r_{qm},\ r_{unif},\ r_{far},\ r_{q_iq_j} & Field--marker, uniform-displacement, far-angle and pairwise cross-moment correlation coefficients. \\
G,\ \widehat G_q,\ G_q & Trial scalar gain, its least-squares estimate and coherent displacement gain of observable $q$ relative to the marker. \\
\bm A(t),\ S_1 & Standardized four-moment amplitude matrix and its leading principal-component variance fraction. \\
\lambda_j^{\mathrm{POD}},\ E_1,\ C_{10},\ N_{90} & Ordered POD eigenvalues, leading POD energy fraction, cumulative energy through ten modes and number of modes required for 90\% variance. \\
A_{ref},\ U_{90} & Reference injected displacement amplitude and amplitude whose lower Wilson bound reaches 90\% detection. \\
p_{stat},\ \Delta\xi & Significance probability in stationarity tests and imposed translation-template offset. \\
R^2,\ \sigma_{a_q},\ \sigma_{a_m} & Weighted coefficient of determination and standard deviations of field-equivalent and marker amplitudes. \\
\bottomrule
\end{tabularx}

\medskip\noindent\textit{Time-scale, stochastic and spectral quantities.}\par\nobreak\vspace{2pt}
\noindent\begin{tabularx}{\textwidth}{@{}>{$}l<{$}X@{}}
\toprule
\text{Symbol} & \text{Definition} \\
\midrule
t_\delta^*=\delta_{10\text{--}90}/D,\ t_s^*=s_{50}/D & Freestream-normalized geometric width and standoff time units. \\
t_{\mathrm{res}} & Raywise residence time, $\int_{s_a}^{s_b}\mathrm ds/|u_n(s)|$. \\
f(t^*),\ \kappa^*,\ \tau_a^*=1/\kappa^* & Effective broadband forcing, dimensionless restoring rate and reduced-model relaxation time. \\
\omega^*,\ f^* & Dimensionless angular frequency and dimensionless cyclic frequency. \\
S_{aa},\ S_{ff} & Displacement power spectral density and white-forcing spectral level. \\
S_{\mathrm{AR1}},\ Q_{\mathrm{AR}} & Sampled AR(1) power spectral density and innovation variance. \\
K & Number of Welch segments in the frequency-domain diagnostic. \\
\mathrm i & Imaginary unit, $\sqrt{-1}$. \\
\mathcal L_{\mathrm{kin}},\ \eta(t) & Stable linearized kinetic evolution operator and stochastic forcing. \\
\bm C,\ \bm Q_\eta,\ \dagger & Stationary state covariance, forcing covariance and adjoint operator. \\
\bottomrule
\end{tabularx}

\medskip\noindent\textit{Operators, subscripts and abbreviations.}\par\nobreak\vspace{2pt}
\noindent\begin{tabularx}{\textwidth}{@{}lX@{}}
\toprule
Notation & Definition \\
\midrule
$\overline{(\cdot)}$, $(\cdot)'$, $(\cdot)^*$ & Temporal mean, fluctuation about that mean and dimensionless quantity. \\
$\operatorname{cov}$, $\operatorname{var}$, $\operatorname{corr}$, $\operatorname{tr}$, $\sum_t$ & Covariance, variance, Pearson correlation, matrix trace and summation over sampled times. \\
Subscripts $\infty$, $w$, $up$, $down$ & Freestream, wall, upstream plateau and downstream plateau. \\
Subscripts $p$, $n$, $m$, $q$, $tr$ & Persistent component, sampling-noise component, marker, field observable and translational mode. \\
AR(1) & First-order autoregressive process. \\
BGK & Bhatnagar--Gross--Krook collision model. \\
DMD, POD, SPOD & Dynamic, proper orthogonal and spectral proper orthogonal decomposition. \\
DSMC & Direct simulation Monte Carlo. \\
kLST & Kinetic linear stability theory. \\
LOgSO-CV & Leave-one-group-size-out cross-validation. \\
PC1, PSD & First principal component and power spectral density. \\
VHS & Variable-hard-sphere molecular collision model. \\
\bottomrule
\end{tabularx}
\endgroup
\section{Introduction}
Hypersonic flow over a blunt body produces a detached compression layer in which momentum redirection, translational heating, internal-energy relaxation and gas--surface interaction occur over coupled spatial scales. When the molecular mean free path is no longer negligible relative to the body dimension, the layer has finite kinetic thickness and the continuum notion of a discontinuous shock surface becomes operational rather than exact. Direct simulation Monte Carlo (DSMC) remains the reference particle method in this regime \citep{Bird1994,Lofthouse2007}, and the circular cylinder remains a canonical geometry because it contains a curved detached front, a stagnation region and strong wall interaction without geometric complexity. Shock extraction in such fields is itself non-trivial: gradient- and schlieren-based definitions agree when a localized front exists, but become diagnostic rather than unique as the layer becomes diffuse \citep{Akhlaghi2017,Akhlaghi2021}.

Shock motion has a long history outside rarefied-gas dynamics. \citet{Plotkin1975} represented a shock as a linearly damped displacement driven by broadband upstream fluctuations, yielding a first-order low-pass response. Experiments and simulations of separated shock-wave/boundary-layer interactions subsequently developed this picture and identified both restoring dynamics and low-frequency forcing associated with the separated region \citep{PoggieSmits2001,PoggieSmits2005,Dussauge2006,Dupont2006,Piponniau2009,TouberSandham2011,PriebeMartin2012}. The modern review of \citet{ClemensNarayanaswamy2014} emphasizes that several mechanisms can generate low-frequency shock motion and that a measured spectrum alone does not identify the forcing. These studies concern turbulent or separated continuum flows, not a rarefied bow shock. Their relevance here is therefore structural rather than literal: they establish a fluid-mechanical question that is sharper than mode detection alone, does the shock layer behave as a damped, spatially organized filter, and what can be inferred about its relaxation time and forcing?

The forcing environment is fundamentally different in DSMC. Molecular and simulator-particle fluctuations are intrinsic to particle fields, and their sampling error depends on particle number, averaging time and the observable being estimated \citep{Stefanov2000,Hadjiconstantinou2003}. Fluctuating-hydrodynamic formulations make explicit that thermal forcing can generate temporally and spatially correlated macroscopic fluctuations \citep{BellGarciaWilliams2007,WilliamsBellGarcia2008,Bell2022}. Molecular simulations further show that stochastic microscopic forcing can seed or alter organized hydrodynamic dynamics, including instability growth and symmetry breaking \citep{Gallis2016,Gallis2017,Gallis2021}. Consequently, the presence of stochastic forcing does not make an observed slow response unphysical, but its amplitude cannot be interpreted independently of simulator-particle weight. The physically transferable quantities are instead the response shape, relaxation time and transfer behaviour, provided they survive changes in sampling and numerical realization.

The specific question here is whether the detached bow layer possesses such a collective displacement coordinate. Let $q(s,\theta,t)$ denote any macroscopic field, $\mean q(s,\theta)$ its temporal mean, $q'=q-\mean q$ its fluctuation, $s$ wall-normal distance along a body-normal ray, $\theta$ polar angle, $t$ dimensional time, and $a(\theta,t)$ the local front-normal displacement amplitude. Translation of a thin or moderately diffuse front then produces the first-order disturbance
\begin{equation}
 q'(s,\theta,t)\simeq-a(\theta,t)\frac{\partial \mean q}{\partial s},
 \label{eq:translation}
\end{equation}
If the layer moves collectively, $a(\theta,t)$ should contain a slowly varying component shared by distant portions of the bow. In DSMC, however, the observed marker also contains finite-particle noise, and the noise is temporally correlated because particle states and sampling accumulators persist between neighbouring outputs. The challenge is therefore to distinguish evidence for a low-pass response from correlated feature error rather than to ask whether the instantaneous field is low rank.

Registration and modal analysis introduce additional pitfalls. Registration methods can remove transport-dominated variability and improve reduced-order compactness \citep{RowleyMarsden2000,Reiss2018,Taddei2020}, but their validity depends on the map and its physical support. A shock-attached grid that crosses the solid body can be filled by repeated wall-adjacent values and generate artificial low rank. Proper orthogonal decomposition (POD) identifies variance-optimal structures \citep{Lumley1967,Sirovich1987,Berkooz1993,Taira2017}; dynamic mode decomposition (DMD) and spectral proper orthogonal decomposition (SPOD) target temporal or frequency-dependent organization \citep{Schmid2010,Towne2018,SchmidtColonius2020}. Power spectra estimated by Welch averaging are useful diagnostics \citep{Welch1967}, but none of these decompositions alone separates persistent covariance from particle-sampling covariance. The present analysis therefore treats modal tools as diagnostics and bases the resolved-component classification on coarse-graining response, prediction across averaging design scales, long-range angular correlation, synthetic controls, completed repeat runs and full-field reconstruction.

Kinetic studies provide the closest rarefied-flow context. DSMC has revealed low-frequency molecular fluctuations inside one-dimensional shocks and analytical models have reproduced part of their macroscopic structure \citep{Sawant2021,Sawant2022}. Most closely, \citet{Senkardesler2026} reported DSMC bow-shock pulsation for Mach-4 flow over a micrometre-scale cylinder using probe spectra and modal diagnostics. That study establishes that low-frequency bow-shock unsteadiness in DSMC is not, by itself, a new observation. The present question is different: whether correlated sampling covariance can be separated from a spatially collective angular displacement, and whether that coordinate survives full-field, independent-seed and particle-loading checks without a discrete spectral line. Kinetic base states have also been combined with linear stability analysis in hypersonic boundary layers and rarefied shear flows \citep{Klothakis2022,Zou2023}, while time-resolved DSMC and data-driven decompositions have exposed organized dynamics in strongly separated high-speed configurations \citep{Karpuzcu2025}. More recently, \citet{Karpuzcu2026} developed a kinetic linear stability theory (kLST) framework and linearized a Boltzmann equation closed with the Bhatnagar--Gross--Krook (BGK) model about one-dimensional normal shocks and showed that non-Maxwellian velocity distributions move spectral branches towards less stable locations. That analysis concerns deterministic perturbations of planar argon shocks, whereas the present work concerns stochastic fluctuations of a curved, wall-coupled nitrogen bow layer. It nevertheless supports a key interpretation: a stable kinetic operator can possess a slowly relaxing macroscopic projection that is continually excited by stochastic forcing.

A recent study used the canonical cylinder geometry for neural-operator development \citep{RoohiShojaAzghadi2026PoF}. That work established the availability and surrogate learnability of cross-regime steady fields; it did not address temporal shock motion or separate persistent covariance from DSMC sampling covariance. The companion study of this cylinder family \citep{RoohiShoja2026Mean} examines mean standoff, mean 10--90 thickness, variable-specific relaxation lengths and parameter-indexed shock-attached POD. Its reported layer broadening is therefore context for the present test, not a novelty claim of this paper. The present paper uses consecutive nitrogen outputs at fixed operating conditions and makes four distinct contributions. First, it audits shock-attached support and quantifies false low-rank structures created by solid-side interpolation. Second, it performs temporal coarse graining before marker extraction and separates persistent and correlated sampling covariance using a calibrated penalized composite fit, block resampling and synthetic controls \citep{Kunsch1989,PolitisRomano1994,Higham2002}. Third, it identifies a common long-range angular displacement shape in the separately analysed $\KnD=0.01$ and $0.025$ records and verifies its shape and memory with independent random seeds and particle loadings. Fourth, it performs complementary checks in density, Mach number, pressure and translational temperature through the translation template in \cref{eq:translation}. 
\section{Configuration and analysis}
\subsection{Direct simulation Monte Carlo data and nondimensionalization}
The configuration is two-dimensional Mach-10 nitrogen flow over a circular cylinder of diameter $D=\SI{0.3048}{m}$ and radius $R=D/2$. The freestream Mach number, speed and temperature are denoted by $M_\infty$, $U_\infty$ and $T_\infty$, respectively, with $M_\infty=10$ and $T_\infty=\SI{200}{K}$; the wall temperature is $T_w=\SI{500}{K}$. Gas--gas collisions use the variable-hard-sphere (VHS) model with nitrogen molecular mass $4.65\times10^{-26}\,\mathrm{kg}$, reference diameter $4.17\times10^{-10}\,\mathrm{m}$ at $\SI{273}{K}$ and viscosity exponent $0.74$. Rotational relaxation is treated with Bird's DS2V program, a two-dimensional direct simulation Monte Carlo solver with internal-energy nonequilibrium models. The cylinder wall is fully diffuse at $T_w$, and the production discretization uses 194 by 100 base divisions with adaptive collision cells targeting approximately 20 simulator particles per cell; the reference loading contains approximately $1.5\times10^6$ simulator particles. Vibrational excitation and chemistry are not included, so the gas should be interpreted as rotationally relaxing nitrogen rather than as a complete thermochemical representation. A calorically perfect normal-shock estimate with specific-heat ratio $\gamma=1.4$ gives an estimated post-shock temperature $T_2\simeq\SI{4.1e3}{K}$, comparable with the nitrogen characteristic vibrational temperature $\theta_v\simeq\SI{3371}{K}$. Let $c_{v,v}$ denote the vibrational contribution to the constant-volume specific heat, $R_s$ the nitrogen specific gas constant, $T$ the equilibrium temperature used in the estimate, and $x=\theta_v/T$ the dimensionless vibrational-temperature ratio. The equilibrium harmonic-oscillator estimate is
\begin{equation}
 \frac{c_{v,v}}{R_s}=\frac{x^2e^x}{(e^x-1)^2},\qquad x=\frac{\theta_v}{T}.
\end{equation}
It gives $c_{v,v}/R_s\simeq0.95$ at $T=T_2$. Vibrational energy could therefore introduce an additional relaxation coordinate in real nitrogen \citep{MillikanWhite1963,Park1988,Bertolotti1998}; the present conclusions are restricted to the rotational model used by DS2V. Let $\lambda_\infty$ denote the freestream mean free path. The diameter-based Knudsen number is
\begin{equation}
 \KnD=\frac{\lambda_\infty}{D}.
\end{equation} The dimensionless convective time is $\tstar=tU_\infty/D$, where $t$ is dimensional time and $U_\infty$ is the freestream speed. The dimensionless centre-to-centre output cadence is denoted by $\Delta\tstar=\Delta t\,U_\infty/D$, where $\Delta t$ is the dimensional interval between output-block centres.

The geometry, molecular model and statistically converged mean fields are shared with the companion mean-flow paper \citep{RoohiShoja2026Mean}, but the data object analysed here is different. The companion work treats one converged field per operating condition and asks how the parameterized mean family collapses. Here each operating condition supplies a consecutive temporal record, and no parameter-to-parameter POD mode is interpreted dynamically. This separation is important because a compact parameterized mean family may coexist with high-dimensional instantaneous fluctuations, while a weak coherent coordinate can remain invisible to energy-ranked field POD.

For the time-resolved campaign, each stored macroscopic field is a short block average accumulated after the preceding output reset; the accumulator is reset after every stored field. The files are therefore not overlapping moving averages. Adjacent blocks remain correlated because the simulator-particle state is continuous across outputs, which is why the sampling component is not assumed white. The analysis uses the measured block-centre time from the output logs rather than treating the file index as physical time.

Nine Knudsen numbers are considered. Every case contains at least 200 consecutive macroscopic fields; the longer records are used for stationarity, covariance and power analyses, whereas exactly 200 snapshots are used for direct field-POD comparison. Let $N_s$ denote the number of stored snapshots. The output cadence is not identical because the datasets were produced in staged campaigns. Every autoregressive and spectral quantity is therefore evaluated with the measured $\Delta\tstar$ of its own record. Quantitative memory comparisons are restricted to the two resolved near-continuum cases, whose cadences are similar ($0.288$ and $0.240$); the coarser high-Knudsen records are used only for detectability bounds and are not used to infer a common frequency trend. On a body-normal ray, $s_{50}$ denotes the density half-jump location and $\delta_{10\text{--}90}$ the distance between the 10\% and 90\% density-transition levels; their extraction is defined below. \Cref{tab:data} summarizes the records and the sector medians of these two quantities.

\begin{table}
\centering
\caption{Time-resolved nitrogen DSMC data. $N_s$ is the number of available snapshots; $\Delta \tstar$ is the centre-to-centre output cadence. The final two columns are sector medians of the density half-jump location and 10--90 width.}
\label{tab:data}
\small
\begin{tabular}{ccccc}
\toprule
$\KnD$ & $N_s$ & $\Delta \tstar$ & $\scenter/R$ & $\deltaten/R$ \\
\midrule
0.010 & 200 & 0.288 & 0.462 & 0.123 \\
0.025 & 600 & 0.240 & 0.468 & 0.224 \\
0.050 & 419 & 0.431 & 0.494 & 0.359 \\
0.075 & 674 & 0.491 & 0.519 & 0.470 \\
0.100 & 400 & 0.369 & 0.538 & 0.561 \\
0.150 & 703 & 0.389 & 0.565 & 0.710 \\
0.250 & 462 & 0.443 & 0.601 & 0.912 \\
0.500 & 350 & 0.508 & 0.661 & 1.243 \\
1.000 & 283 & 0.789 & 0.756 & 1.734 \\
\bottomrule
\end{tabular}
\end{table}

\subsection{Half-jump marker and physical shock-attached support}
\begin{figure}[t]
\centering
\includegraphics[width=\textwidth]{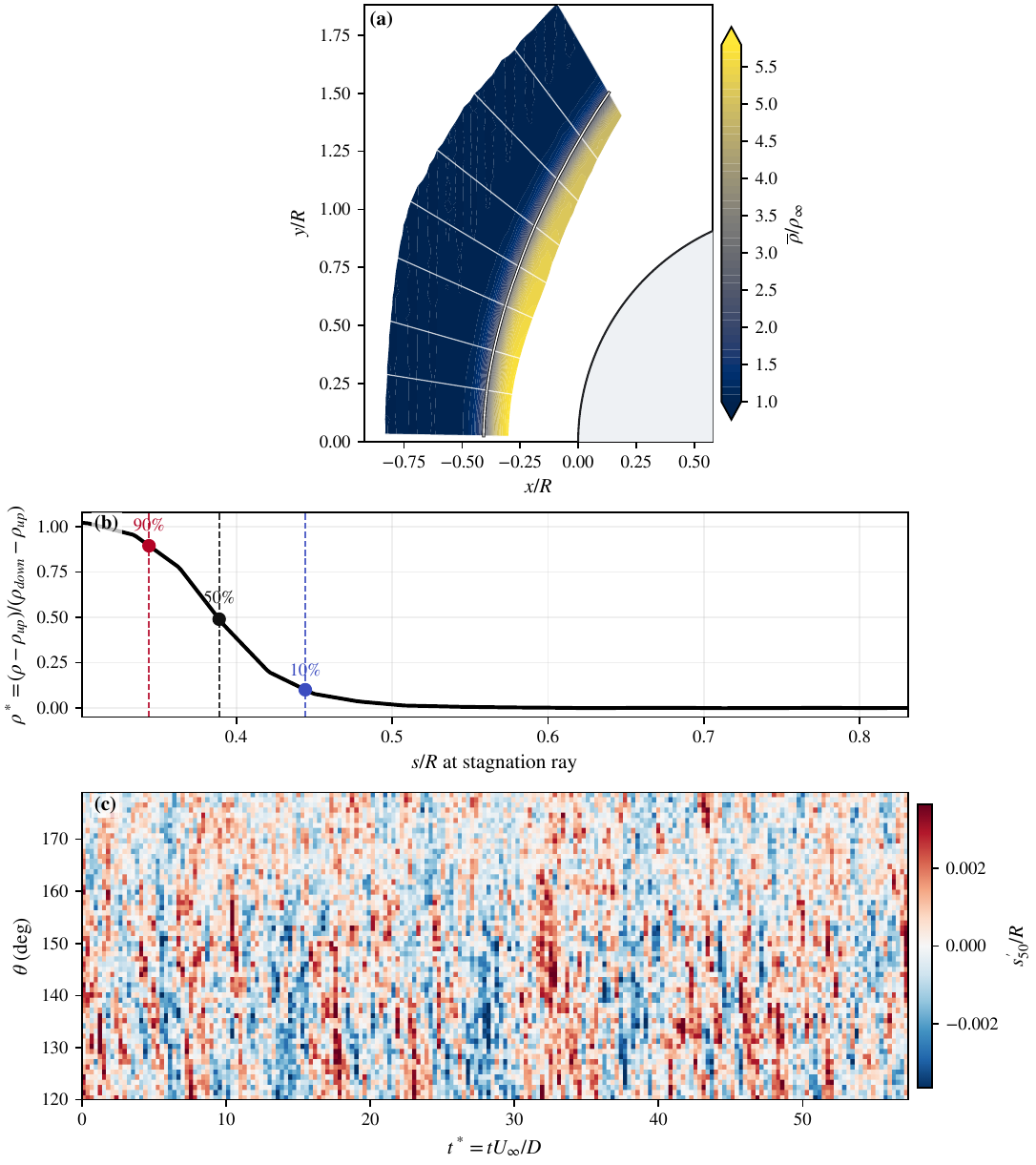}
\caption{Shock-attached extraction at $\KnD=0.01$, shown as three vertically stacked panels for readability. (a) Mean normalized density, selected body-normal rays and the density half-jump centre; only gas-side support is sampled. (b) Stagnation-ray normalized density profile with the 10, 50 and 90\% transition markers. (c) Centred time-resolved half-jump location over the upstream analysis sector.}
\label{fig:geometry}
\end{figure}

The upper upstream sector is represented by 60 body-normal rays over $120^\circ\leq\theta\leq179^\circ$. In the polar convention used here the upstream stagnation line is at $\theta=180^\circ$. On each ray, $r$ is distance from the cylinder centre and $s=r-R$ is distance from the cylinder wall. Let $\rho(s,\theta,t)$ be the instantaneous density and let $\rho_{up}(\theta,t)$ and $\rho_{down}(\theta,t)$ denote robust local upstream and downstream density plateaux. They define the normalized transition
\begin{equation}
 \rho^*(s,\theta,t)=\frac{\rho-\rho_{up}}{\rho_{down}-\rho_{up}}.
\end{equation}
The centre $\scenter(\theta,t)$ is the $\rho^*=0.5$ crossing and $\deltaten(\theta,t)$ is the distance between the 0.1 and 0.9 crossings. The attached coordinate is
\begin{equation}
 \xi=\frac{s-\scenter}{\deltaten}, \qquad -1\leq\xi\leq4.
 \label{eq:xi}
\end{equation}
Each body-normal ray is evaluated on 900 raw radial samples extending to $s/R=8$. The plateau and crossing search excludes $s/R<0.25$, and the density profile on each ray is smoothed with a Gaussian kernel of standard deviation four raw samples before the 10, 50 and 90\% crossings are interpolated. The common attached grid contains 260 points in $\xi$. A point is retained only when $s/R\geq0.02$; unsupported points are masked rather than filled from the nearest gas cell. The raw crossing exclusion and the attached-field support mask serve different purposes: the former protects plateau estimation near the wall, whereas the latter prevents solid-side values from entering field statistics. \Cref{fig:geometry} illustrates the map, the 10--50--90 construction and a time-resolved marker field.

\begin{figure}[t]
\centering
\includegraphics[width=\textwidth]{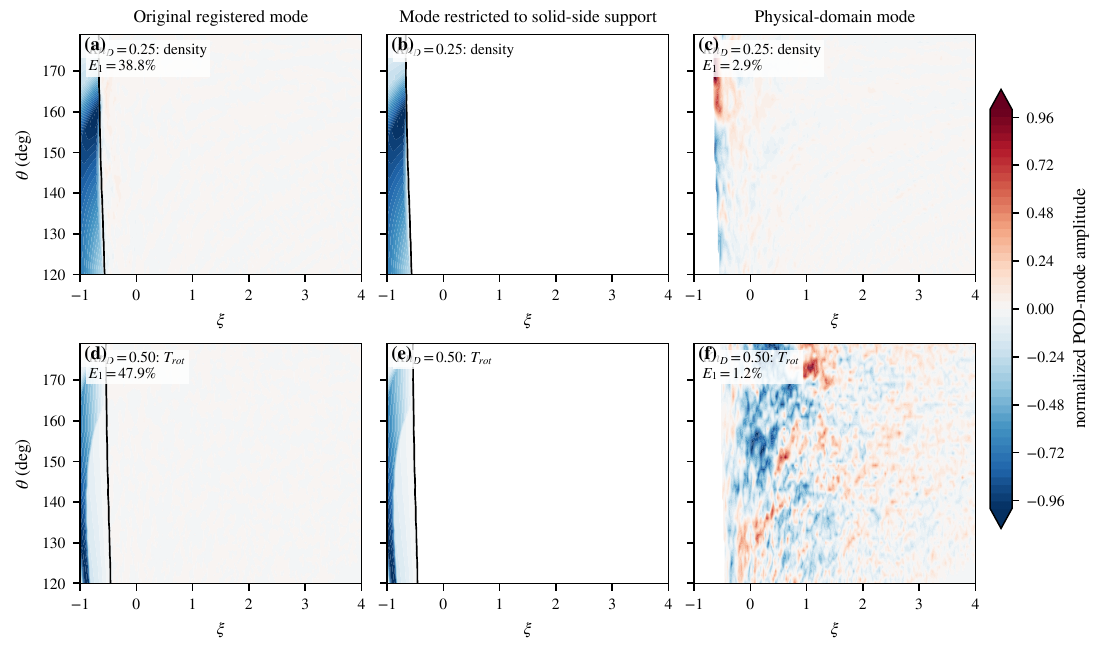}
\caption{Physical-support audit of the apparent low-rank structures. The left column shows leading modes from a registration that admitted solid-side points, the middle column isolates their non-physical contribution, and the right column shows the physical-domain result. The black curve in the original maps marks the $s/R=0.02$ support boundary. The collapse of $E_1$ demonstrates that solid-side remapping can create false coherence.}
\label{fig:support}
\end{figure}

The physical-support restriction is essential rather than cosmetic. A preliminary registration route allowed portions of the attached window to lie inside the cylinder and filled them by nearest-neighbour remapping. As shown in \cref{fig:support}, the apparent leading density energy at $\KnD=0.25$ decreased from 38.8\% to 2.9\% after clipping; the corresponding rotational-temperature energy at $\KnD=0.5$ decreased from 47.9\% to 1.2\%. The original modes are concentrated at the solid-side edge of the attached coordinate. Here and in \cref{fig:support}, $E_1$ denotes the leading proper-orthogonal-decomposition energy fraction, i.e. the leading POD eigenvalue divided by the sum of all POD eigenvalues. We retain this failed route as a negative control and use only physical-domain results below.

\subsection{Temporal coarse graining and correlated-noise covariance}
To test whether marker fluctuations survive increased sampling, groups of $m=1,2,4,8,16$ consecutive density fields are averaged \emph{before} re-extracting $\scenter$ and $\deltaten$. Let $C_m$ denote the angular covariance matrix of the centre marker obtained after averaging groups of $m$ consecutive fields. For a first-order autoregressive [AR(1)] process, let $\phi$ be its lag-one correlation coefficient, let $k$ denote integer lag within the averaging block, and let $A_m(\phi)$ denote the resulting variance-attenuation factor. Under block averaging,
\begin{equation}
 A_m(\phi)=\frac{1}{m^2}\left[m+2\sum_{k=1}^{m-1}(m-k)\phi^k\right].
 \label{eq:attenuation}
\end{equation}
Let $C_p$ and $C_n$ denote the persistent and correlated sampling covariance matrices, respectively, and let $\phi_p$ and $\phi_n$ denote their corresponding lag-one AR(1) coefficients. We model the measured covariance as
\begin{equation}
 C_m=A_m(\phi_p)C_p+A_m(\phi_n)C_n.
 \label{eq:covmodel}
\end{equation} Equation~\eqref{eq:attenuation} is exact for a linearly averaged marker series. Here fields are averaged first and a nonlinear crossing is re-extracted, so its use in \cref{eq:covmodel} is a first-order attenuation approximation rather than an identity. The 10--90 width channel supplies a diagnostic estimate of $\phi_n$ because it is especially sensitive to local sampling fluctuations; the centre channel then determines $\phi_p$, $C_p$ and $C_n$. At $\KnD=0.01$ the width-based estimate $\phi_n=-0.25$ reaches the lower search bound, and the pointwise and angular-mean width attenuation cannot be represented by one AR(1) coefficient. We therefore treat $\phi_n$ as a boundary calibration, not a measured physical parameter, repeat the inference over the sensitivity range reported in Appendix~C, and require the independent-seed and particle-loading controls below. The fitted covariance matrices are constrained to be positive semidefinite.

The fit combines normalized squared errors for five group sizes in the pointwise trace and angular-mean variance with the mean squared autocorrelation error at eight lags. If $J$ is this composite objective, $n_c=18$ the number of retained summary terms and $k$ the number of fitted scalar parameters, we define the small-sample penalized composite score
\begin{equation}
 IC_c=n_c\log J+2k+\frac{2k(k+1)}{n_c-k-1},
 \label{eq:icc}
\end{equation}
and $\Delta IC_c=IC_{c,\mathrm{noise}}-IC_{c,\mathrm{two}}$. Because the 18 summaries are dependent, heteroscedastic and obtained from one trajectory, $IC_c$ is not a likelihood-based Akaike information criterion and its magnitude has no standard Akaike interpretation. It is used only as one calibrated composite-fit diagnostic. Leave-one-group-size-out cross-validation (LOgSO-CV) omits one averaging design scale at a time; it tests interpolation across group size, not generalization to an independent trajectory. The reported LOgSO-CV ratio is the two-component error divided by the noise-only error, so a ratio below unity favours the two-component representation. Moving-block resampling, far-angle covariance for $|\theta-\theta'|\geq15^\circ$ (where $\theta'$ is the second ray angle), and matched synthetic controls provide additional diagnostics. The dimensionless persistent-memory time scale $\tau_p^*$ is
\begin{equation}
 \tau_p^*=-\frac{\Delta \tstar}{\log \phi_p}.
\end{equation}
A persistent component is classified as resolved only when the two-component model improves design-scale prediction, its block-resampled composite-score preference is positive, long-range covariance is positive, synthetic false detections remain controlled, the positive-semidefinite covariance projection does not require a large correction, and the response shape and memory reproduce in the independent-seed and particle-loading controls.

\subsection{Stationarity, window persistence and control experiments}
A persistent covariance can be inferred only from a record that is sufficiently stationary over the analysis interval. For every full record we test the angular-mean centre and width for a linear trend and compare the variances of the first and second halves using Levene's test. We use $p_{stat}$ to denote the significance probability returned by these statistical tests, distinct from the pressure symbol $p$. The full $\KnD=0.025$ record gives centre trend and variance-stationarity values $p_{stat}=0.407$ and $0.149$, respectively; the corresponding width values are 0.752 and 0.098. The width variance is therefore close to the conventional 5\% threshold but does not reject stationarity. The full records at $\KnD=0.05$, $0.075$ and $0.15$ likewise show no significant centre trend. The $\KnD=0.10$ centre has a weaker margin ($p_{stat}=0.070$), motivating the window-resolved interpretation used below.

Stationarity of low-order statistics does not guarantee persistence of the inferred mode. We therefore repeat the complete covariance fit in overlapping 192-snapshot windows stepped by 48 snapshots. A window passes only if the persistent-plus-sampling model has $\Delta IC_c>10$, improves LOgSO-CV, recovers a mode aligned by at least 0.70 with the $\KnD=0.01$ reference, has positive far-angle covariance and requires less than 20\% relative positive-semidefinite projection correction. The threshold is an empirically stress-tested decision rule for this composite score, not a universal information-theoretic cutoff. Because adjacent windows overlap by 75\%, window pass counts are descriptive and are not treated as independent binomial trials.

Two synthetic controls complete the calibration. The negative control simulates only the fitted correlated sampling covariance and measures false detections. The positive control injects the $\KnD=0.01$ rank-one angular shape and memory and measures the response of a computationally tractable four-gate subset of the final classifier. It therefore calibrates sensitivity to that specified reference, not the power of every final decision gate or robustness to a misspecified mode shape. Injection amplitude is varied from zero to six times the reference displacement standard deviation $A_{ref}$ measured at $\KnD=0.01$, and Wilson intervals are used for the detection fraction. We denote by $U_{90}$ the injected amplitude required for the lower Wilson bound to reach 90\%; this conservative definition differs from interpolation through the point estimates. Percentile block-resampling ranges are reported as properties of the resampling distribution, not as independent-sample confidence intervals.

\subsection{Independent-realization and particle-loading controls}
A controlled $2\times2$ repeat campaign at $\KnD=0.01$ tests dependence on random realization and simulator-particle loading. Let $N_p$ denote the reference loading of $1.5\times10^6$ simulator particles. The four calculations use independently initialized random seeds 104729 and 130363 at both $N_p$ and $2N_p=3.0\times10^6$. The base mesh, geometry, molecular model, wall condition, adaptive-cell policy and analysed output protocol are common to all four runs. Each record contains 600 non-overlapping output blocks with 63 accumulated samples per block. Doubling the loading halves the simulator-particle weight, from $F_N=3.9868798250079\times10^{13}$ to $1.9934400000158\times10^{13}$ real molecules per simulator particle. Each seed begins as a new DS2V calculation; a changed seed is not introduced by continuing a restart from the other realization, and interrupted continuations restore the saved random-number-generator state.

The same frozen inference pipeline is applied to all four runs. Its case-matched calibration uses 400 sampling-only synthetic records per case. With $b$ null statistics at least as large as the observed value, the finite Monte Carlo probability is $(b+1)/(400+1)$; the per-case decision level is 0.01, giving a four-case union-bound familywise level of at most 0.04. Repeat comparisons are evaluated by absolute angular-mode alignment, overlap of block-resampling memory ranges, the ratio of persistent memory times, and the ratios of raw and inferred persistent variances. The results are reported in Appendix~B. They test robustness of classification, shape and memory, not invariance of absolute displacement amplitude with simulator-particle weight.

\subsection{Full-field displacement-template validation}
The marker analysis is tested against the complete macroscopic fields. Here $\rho$ is density, $M$ is local Mach number, $T_{tr}$ is translational temperature and $p$ is pressure. Let $g(\theta)$ denote the normalized physical angular mode inferred at $\KnD=0.01$. For $q\in\{\rho,M,T_{tr},p\}$, the corresponding translation template $\Psi_q$ is
\begin{equation}
 \Psi_q(s,\theta)=-g(\theta)\frac{\partial \mean q}{\partial s}.
 \label{eq:template}
\end{equation}
Before defining the amplitude, let $\langle u,v\rangle_W=\sum_j W_j u_jv_j$ denote the discrete weighted inner product over all retained attached-grid points, where $j$ indexes those points and $W_j$ is the corresponding non-negative weight. The least-squares equivalent displacement amplitude $a_q(t)$ is then
\begin{equation}
 a_q(t)=\frac{\langle q'(t),\Psi_q\rangle_W}{\langle \Psi_q,\Psi_q\rangle_W}.
 \label{eq:amplitude}
\end{equation}
The weight field $W$ is concentrated on the density-gradient core and normalized to give each retained ray equal total weight. This construction uses every supported field point rather than one crossing. Moving-block resampling ranges and circular time-shift tests quantify the correlation between $a_q(t)$ and the separately extracted marker amplitude. These checks are complementary rather than statistically independent: all macroscopic moments are accumulated from the same simulator particles, and the density marker and density template share the same density field. We also form conditional composites from the largest positive and negative marker events and compare them with the translation reconstruction.

\subsection{Slow-memory and multi-moment synchronization diagnostics}
The translation-template amplitudes permit two additional diagnostics that connect the statistical coordinate to flow physics. Let $a_m$ be the scalar angular marker amplitude, $a_q$ the equivalent displacement obtained from the full-field template, $r_{qm}=\operatorname{corr}(a_q,a_m)$ their Pearson correlation coefficient, and $\sigma_{a_q}$ and $\sigma_{a_m}$ their standard deviations. The regression or coherent displacement gain of moment $q$ relative to the marker is
\begin{equation}
 G_q=\frac{\operatorname{cov}(a_q,a_m)}{\operatorname{var}(a_m)}
 =r_{qm}\frac{\sigma_{a_q}}{\sigma_{a_m}}.
 \label{eq:gain}
\end{equation} $G_q=1$ corresponds to a field translating with the same amplitude as the marker; $G_q<1$ indicates partial moment participation, additional relaxation or a reduced signal-to-noise ratio. Unlike the projection-energy fraction, $G_q$ retains the sign and amplitude of the coherent component and therefore provides a direct comparison between compression, pressure, velocity--sound-speed and thermal responses.

Second, the four field amplitudes $a_\rho$, $a_p$, $a_M$ and $a_{T_{tr}}$ are standardized and assembled into a matrix $\bm A(t)$. Let $\lambda_1(\operatorname{cov}\bm A)$ denote the largest eigenvalue of the amplitude covariance matrix and let $\operatorname{tr}(\operatorname{cov}\bm A)$ denote its trace. The leading principal-component fraction is
\begin{equation}
 S_1=\frac{\lambda_1(\operatorname{cov}\bm A)}{\operatorname{tr}(\operatorname{cov}\bm A)}.
 \label{eq:synch}
\end{equation}
Thus $S_1$ measures the degree of multi-moment synchronization. A value near unity means that all four moments are organized predominantly by one common temporal amplitude. We also report the median pairwise moment correlation and the correlation of the first principal component with the marker.

The collective memory is compared with two freestream-normalized geometric time units. Because $\tstar=tU_\infty/D$, division of the mean 10--90 width and mean standoff by $U_\infty$ gives
\begin{equation}
 t_\delta^*=\frac{\deltaten}{D},\qquad
 t_s^*=\frac{\scenter}{D}.
 \label{eq:passage}
\end{equation}
These are nominal geometric comparisons, not gas residence times within the compression layer. An actual normal residence time would require the local raywise velocity,
\begin{equation}
 t_{\mathrm{res}}=\int_{s_a}^{s_b}\frac{\mathrm ds}{|u_n(s)|},
\end{equation}
where $u_n$ is the local body-normal velocity and $s_a,s_b$ delimit the layer. That quantity is not estimated here, so the ratios $\tau_p^*/t_\delta^*$ and $\tau_p^*/t_s^*$ establish separation only from freestream-normalized geometric units and do not by themselves exclude an advective mechanism.

\section{Results}
\subsection{A broadening mean layer with a high-rank stochastic background}\label{sec:results-pod}
\begin{figure}[t]
\centering
\includegraphics[width=\textwidth]{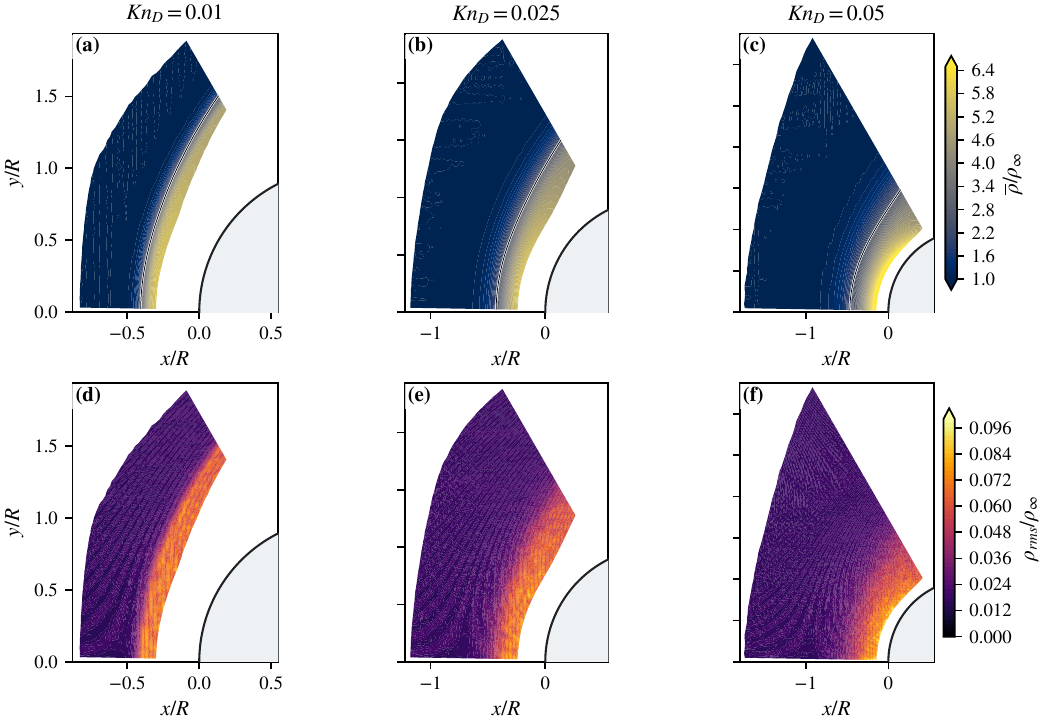}
\caption{Mean compression-layer geometry and raw fluctuation level. Top row: $\mean\rho/\rho_\infty$ with the mean half-jump centre overlaid. Bottom row: $\rms\rho/\rho_\infty$, where $\rho_\infty$ is the freestream density and $\rms\rho$ denotes the density r.m.s. fluctuation. The same colour range is used across cases in each row. The mean layer broadens strongly while the stochastic field remains spatially distributed.}
\label{fig:meanrms}
\end{figure}

The mean density layer broadens continuously over the low-Knudsen interval (\cref{fig:meanrms}). The half-jump centre changes only modestly between $\KnD=0.01$ and $0.025$, but $\deltaten/R$ nearly doubles. At $\KnD=0.05$ the transition spans a substantial fraction of the radius. In contrast, the raw density root-mean-square (r.m.s.) fluctuation remains distributed over the complete compression region and is not concentrated on a single deterministic shape.

\begin{figure}[t]
\centering
\includegraphics[width=\textwidth]{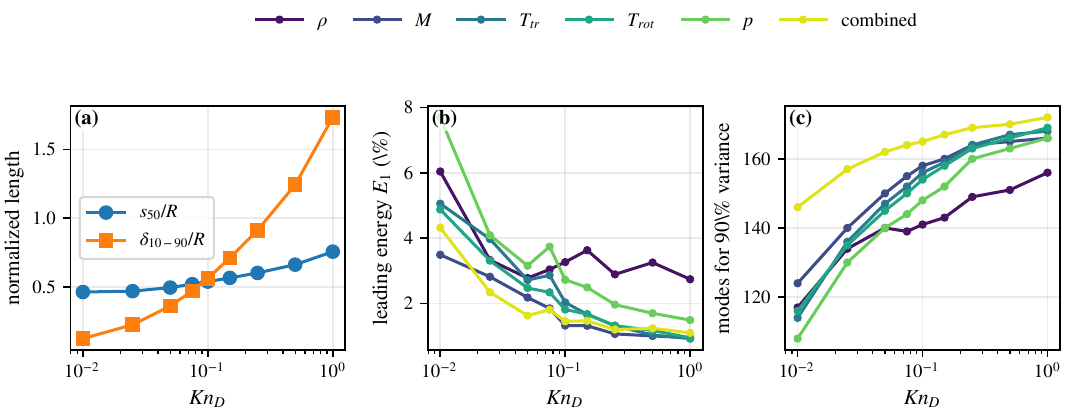}
\caption{Mean geometry and corrected field dimensionality over the nine-case Knudsen sweep. (a) Sector-median half-jump location and 10--90 width. (b) Corrected leading POD energy (POD: proper orthogonal decomposition). (c) Number of modes required for 90\% variance. Exactly 200 snapshots are used at every $\KnD$ for the POD comparison.}
\label{fig:highrank}
\end{figure}

This high dimensionality is confirmed by the corrected physical-domain proper orthogonal decomposition (POD). Let $\lambda_j^{\mathrm{POD}}$ be the POD eigenvalues ordered from largest to smallest. We define the leading energy fraction $E_1=\lambda_1^{\mathrm{POD}}/\sum_j\lambda_j^{\mathrm{POD}}$, the cumulative ten-mode fraction $C_{10}=\sum_{j=1}^{10}\lambda_j^{\mathrm{POD}}/\sum_j\lambda_j^{\mathrm{POD}}$, and $N_{90}$ as the smallest number of modes whose cumulative fraction reaches 90\%. \Cref{fig:highrank} compares the mean geometry with field compactness across all nine Knudsen numbers. For the combined state, $E_1$ decreases from 4.32\% at $\KnD=0.01$ to 1.10\% at $\KnD=1$, and the number of modes required for 90\% variance increases from 146 to 172. Density, pressure, Mach number and both temperatures follow the same qualitative conclusion. The intermediate points do not reveal modal condensation near $\KnD=0.1$; they form a smooth high-rank sequence.

\begin{table}
\centering
\caption{Corrected common-200 POD metrics for the combined density--Mach--temperature--pressure state. $C_{10}$ is cumulative energy through ten modes.}
\label{tab:pod}
\small
\begin{tabular}{cccc}
\toprule
$\KnD$ & $E_1$ (\%) & $C_{10}$ (\%) & $N_{90}$ \\
\midrule
0.010 & 4.32 & 21.91 & 146 \\
0.025 & 2.34 & 14.87 & 157 \\
0.050 & 1.63 & 12.12 & 162 \\
0.075 & 1.81 & 11.84 & 164 \\
0.100 & 1.45 & 11.08 & 165 \\
0.150 & 1.47 & 10.46 & 167 \\
0.250 & 1.20 & 9.26 & 169 \\
0.500 & 1.24 & 8.99 & 170 \\
1.000 & 1.10 & 8.45 & 172 \\
\bottomrule
\end{tabular}
\end{table}

The collective coordinate identified below is therefore not the leading field POD mode. It is a low-amplitude covariance direction embedded in a broadband background. This distinction prevents the common but incorrect inference that a coherent marker necessarily implies a low-rank complete flow field.

\subsection{Coarse graining reveals a persistent low-Knudsen component}
\begin{figure}[t]
\centering
\includegraphics[width=\textwidth]{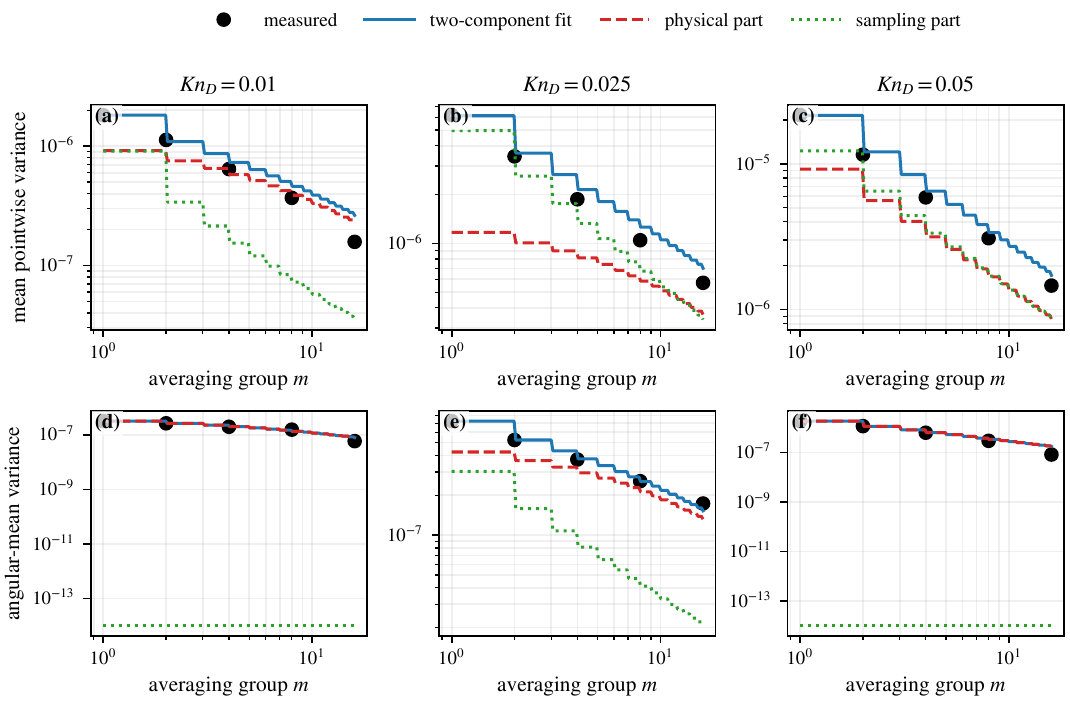}
\caption{Temporal coarse graining of the density half-jump centre. Top row: mean pointwise angular variance; bottom row: variance of the angular-mean displacement. Symbols are measurements after averaging $m$ consecutive fields before marker extraction. Solid curves are the fitted persistent-plus-sampling representation; dashed and dotted curves show its components. The $\KnD=0.05$ decomposition is displayed as a diagnostic but is rejected by design-scale prediction and covariance-projection checks.}
\label{fig:coarse}
\end{figure}

\Cref{fig:coarse} shows how marker variance changes when density fields are averaged before extraction. The pointwise centre variance decreases strongly with $m$ in all cases, confirming that raw marker motion is dominated by finite-particle fluctuations. The angular-mean variance behaves differently. At $\KnD=0.01$ and $0.025$, the fitted persistent contribution attenuates on a finite time scale and remains distinguishable from the sampling component. At $\KnD=0.05$, a nominal two-component fit can be drawn, but it fails the design-scale prediction and positive-semidefinite-projection tests below.

The inference metrics are summarized in \cref{tab:inference}. At $\KnD=0.01$, the two-component representation gives $\Delta IC_c=76.4$, with a block-resampling range of 33.3--73.2, and reduces the leave-one-group-size-out error from 0.784 to 0.432. At $\KnD=0.025$, the 600-snapshot record gives $\Delta IC_c=59.1$ and reduces the error from 0.437 to 0.254. These score separations describe the calibrated composite objective in \cref{eq:icc}; they are not Akaike likelihood evidence. The persistent angular modes have uniform-displacement correlations $r_{unif}=0.868$ and $0.876$, where $r_{unif}$ is the normalized correlation of the leading angular mode with a constant same-signed displacement. They also have positive far-angle correlations $r_{far}$, defined from the normalized covariance at angular separations of at least $15^\circ$, and persistent-memory times $\tau_p^*=0.653$ and $0.724$. Their direct shape correlation is 0.972. The same structure is therefore recovered from two separately analysed rarefaction levels and records of different length and cadence.

\begin{table}
\centering
\caption{Correlated-noise inference for the resolved cases and the $\KnD=0.05$ diagnostic. Brackets denote 2.5--97.5 percentile ranges of the moving-block resampling distribution, not independent-sample confidence intervals. The LOgSO-CV column is the two-component/noise-only design-scale error ratio.}
\label{tab:inference}
\small
\begin{tabular}{ccccc}
\toprule
$\KnD$ & $N_s$ & $\tau_p^*$ [range] & $\Delta IC_c$ [range] & LOgSO-CV ratio \\
\midrule
0.010 & 200 & 0.653 [0.425,0.786] & 76.4 [33.3,73.2] & 0.55 \\
0.025 & 600 & 0.724 [0.239,1.019] & 59.1 [29.0,64.4] & 0.58 \\
0.050 & 419 & 0.280 [0.222,0.611] & 18.0 [1.9,37.5] & 3.74 \\
\bottomrule
\end{tabular}

\medskip
\begin{tabular}{ccc}
\toprule
$\KnD$ & $r_{unif}$ [range] & $r_{far}$ [range] \\
\midrule
0.010 & 0.868 [0.704,0.924] & 0.184 [0.065,0.331] \\
0.025 & 0.876 [0.781,0.891] & 0.267 [0.120,0.287] \\
0.050 & 0.767 [0.241,0.795] & 0.072 [-0.002,0.089] \\
\bottomrule
\end{tabular}
\end{table}

\subsection{The resolved classification and angular shape survive the $2\times2$ controls}
All four repeat calculations pass the case-matched synthetic-null calibration. Their observed penalized composite-score differences span $\Delta IC_c=63.8$--98.7, whereas the corresponding 99th percentiles of the matched sampling-only distributions span only 20.9--33.7. No one of the 400 null records for any case reaches its observed statistic, giving the finite Monte Carlo probability $p_{\mathrm{MC}}=1/401=0.00249$ per case. The two-component/noise-only LOgSO-CV ratios are 0.276--0.394, and the 2.5th-percentile far-angle correlations remain positive (0.047--0.172). Thus the repeat classification is not inherited from the former generic $\Delta IC_c>10$ rule; it is calibrated against the complete statistic under a case-matched null.

The seed-to-seed mode alignments are 0.990 at $N_p$ and 0.977 at $2N_p$. The $N_p$-to-$2N_p$ alignments are 0.967 for seed 104729 and 0.974 for seed 130363, and every pair has overlapping block-resampling ranges for $\tau_p^*$. Density field--marker correlations remain 0.946--0.954 and pressure correlations 0.903--0.918 across the four calculations. These results support realization and loading robustness of the resolved classification, normalized angular geometry and memory.

The amplitude does not show the same invariance. On doubling the particle loading, the raw marker-variance ratios are 0.457 and 0.479, while the inferred persistent-variance ratios are 0.491 and 0.517. The latter remain close to the inverse-loading factor. Consequently, the control establishes reproducibility of shape and memory but does not establish a particle-number-independent absolute fluctuation variance. The complete per-case and pairwise audit is given in \cref{tab:controlcases,tab:controlpairs}.

\subsection{Angular covariance and collective displacement shape}
\begin{figure}[t]
\centering
\includegraphics[width=\textwidth]{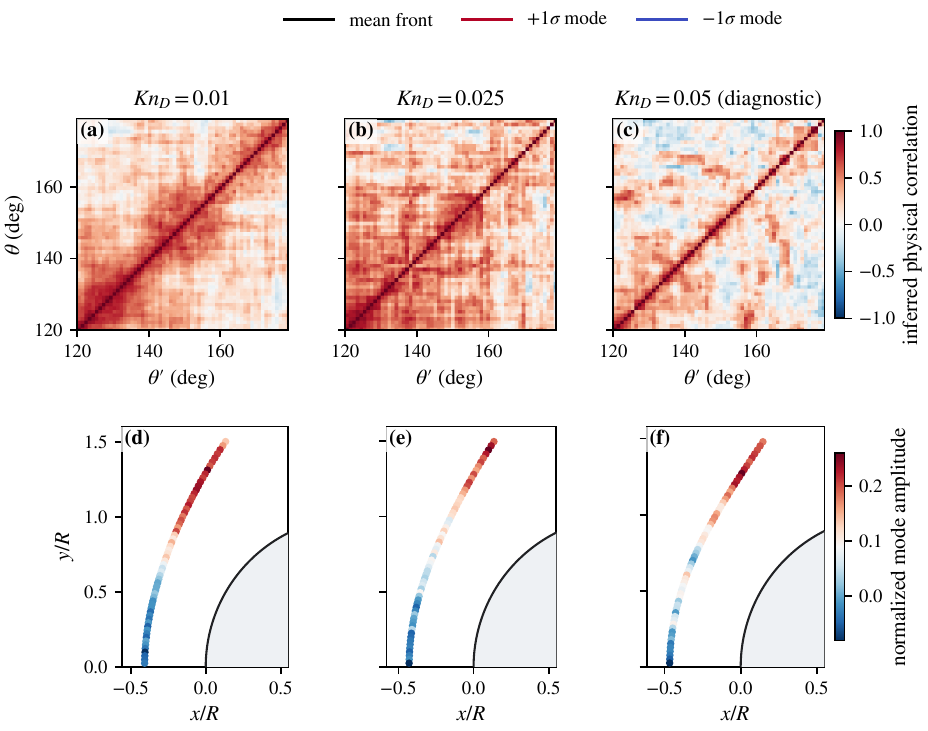}
\caption{Noise-separated angular covariance and inferred displacement shape. Top row: correlation matrices of $C_p$. Bottom row: mean density half-jump front, normalized mode amplitude and $\pm1\sigma_1$ leading-mode displacements, where $\lambda_1(C_p)$ is the largest eigenvalue of the persistent covariance matrix $C_p$ and $\sigma_1=\sqrt{\lambda_1(C_p)}$ is one standard deviation of the leading covariance coordinate. The $\KnD=0.05$ result is diagnostic and is not accepted as a resolved persistent mode.}
\label{fig:covariance}
\end{figure}

The persistent covariance matrices in \cref{fig:covariance} provide the clearest visual evidence. At $\KnD=0.01$ and $0.025$, broad positive off-diagonal regions connect rays separated by tens of degrees. Their leading modes are same-signed over the complete sector and displace the mean half-jump front with a non-uniform angular envelope. The mode should therefore be described as a collective displacement, not as rigid translation. At $\KnD=0.05$, the matrix is patchy, the relative positive-semidefinite projection correction is 54\%, and the two-component LOgSO-CV error is 3.7 times the noise-only error. That panel is retained only to show the failure of identifiability.

\begin{figure}[t]
\centering
\includegraphics[width=\textwidth]{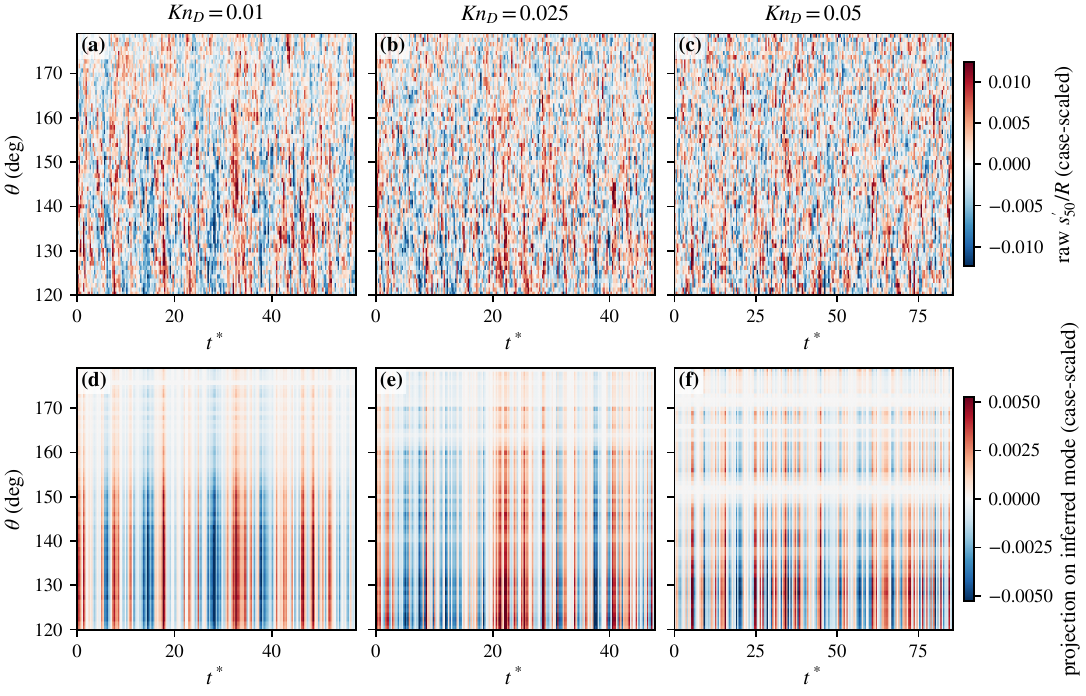}
\caption{Angular-time displacement maps. Top row: raw centred half-jump marker. Bottom row: projection onto the inferred leading angular mode. Each case is plotted with its own symmetric colour scale to expose structure; the panels are not direct amplitude comparisons. Quantitative model selection is required because a visually organized projection can exist even when the covariance decomposition fails.}
\label{fig:spacetime}
\end{figure}

The angular-time maps in \cref{fig:spacetime} show why a direct visual interpretation of raw markers is unreliable. All three cases contain localized streaks and speckle. Projection onto the inferred mode exposes a sector-wide component in the two resolved cases, but a visually structured projection can also be constructed in the rejected case. The combined covariance and repeat-run checks, rather than appearance alone, determine whether the projection is retained.

The inferred root-mean-square collective displacement is small, approximately $5.6\times10^{-4}R$ and $6.5\times10^{-4}R$ at $\KnD=0.01$ and $0.025$, or about 0.45\% and 0.29\% of the corresponding 10--90 width. Moreover, $|g(\theta)|$ correlates with the inverse maximum mean density gradient by approximately 0.84 and 0.94. Because crossing-position uncertainty scales as $|\partial_s\mean\rho|^{-1}$, the angular envelope alone cannot rule out marker sensitivity. The full-field zero-shift localization, multi-moment consistency and completed seed/loading repeats reduce this ambiguity, but invariance to a distinct pressure- or gradient-based front marker is not established by the present archive and remains a stated limitation.

\subsection{The same coordinate is present in the complete macroscopic fields}
\begin{figure}[t]
\centering
\includegraphics[width=\textwidth]{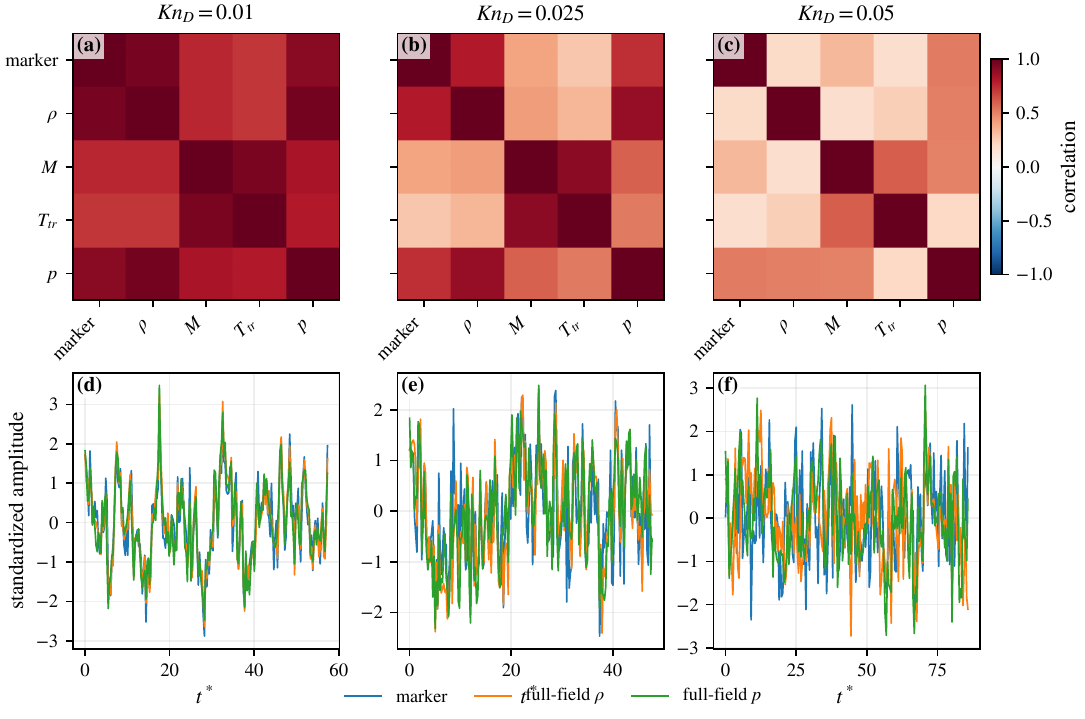}
\caption{Complementary full-field check. Top row: correlation matrices of the marker and translation-template amplitudes extracted from density, Mach number, translational temperature and pressure. Bottom row: standardized marker, density and pressure amplitudes. The observables share simulator particles and are therefore not statistically independent; the resolved cases nevertheless show a common multi-moment time dependence, while the $\KnD=0.05$ case is substantially weaker.}
\label{fig:multimoment}
\end{figure}

The full-field matched filter tests whether the ray marker corresponds to translation of the entire compression layer. \Cref{fig:multimoment} shows amplitude correlations and representative time series. At $\KnD=0.01$, density and pressure amplitudes correlate with the marker by 0.946 and 0.901. Mach number and translational temperature give 0.755 and 0.707. A principal component formed from the four field amplitudes contains 87.3\% of their variance and correlates 0.886 with the marker. At $\KnD=0.025$, density and pressure remain strong at 0.790 and 0.727; the four-moment consensus contains 70.6\% and correlates 0.647 with the marker. At $\KnD=0.05$, the density correlation falls to 0.183 and cross-variable consensus weakens.

\begin{table}
\centering
\caption{Full-field displacement-template metrics. Each entry is marker correlation followed by the median fraction of raw instantaneous field variance aligned with the template. Correlation intervals are given in the text.}
\label{tab:template}
\small
\begin{tabular}{ccccc}
\toprule
$\KnD$ & $\rho$ & $M$ & $T_{tr}$ & $p$ \\
\midrule
0.010 & 0.946 / 5.29\% & 0.755 / 2.30\% & 0.707 / 1.24\% & 0.901 / 4.03\% \\
0.025 & 0.790 / 2.37\% & 0.395 / 1.73\% & 0.273 / 0.97\% & 0.727 / 1.60\% \\
0.050 & 0.183 / 0.67\% & 0.325 / 1.06\% & 0.164 / 0.46\% & 0.510 / 0.77\% \\
\bottomrule
\end{tabular}
\end{table}

The translation coordinate is weak in an energetic sense. Even at $\KnD=0.01$, the median projected fraction is 5.3\% for density and 4.0\% for pressure; it is smaller for the thermal and Mach-number fields. This is not a defect of the result. It demonstrates that the coherent motion coexists with high-dimensional kinetic fluctuations and explains why conventional field POD does not isolate it as a dominant mode.

The full-field agreement is not a purely algebraic consequence of using density to define the marker, but neither is it an independent experiment. Pressure, Mach number and translational temperature are projected separately using their own mean gradients, and their temporal amplitudes are not used in marker extraction; however, all moments share simulator particles and the density template shares the density record with the marker. Circularly shifting the marker relative to each field gives null-test probabilities of 0.0054 for all four moments at $\KnD=0.01$ and for density, Mach number and pressure at $\KnD=0.025$; the density probability at $\KnD=0.05$ rises to 0.0162 and its resampled correlation range approaches zero. The spatial template is also shifted by $\Delta\xi=\{-1,-0.5,0.5,1\}$, where $\Delta\xi$ is its offset from the physical zero-shift position. At $\KnD=0.01$, the unshifted density template captures 5.29\% of variance, compared with 1.25--2.39\% for upstream shifts and 0.29--1.99\% for downstream shifts. At $\KnD=0.025$, the unshifted density value is 2.37\%, approximately twice the $\Delta\xi=-0.5$ value. Pressure exhibits the same zero-shift maximum in both resolved cases. At $\KnD=0.05$, by contrast, the density projection is larger for $\Delta\xi=0.5$ and 1 than at zero shift. The resolved low-Knudsen signal is therefore localized on the actual mean compression gradient, whereas the diagnostic $0.05$ response is not.

\begin{figure}[t]
\centering
\includegraphics[width=\textwidth]{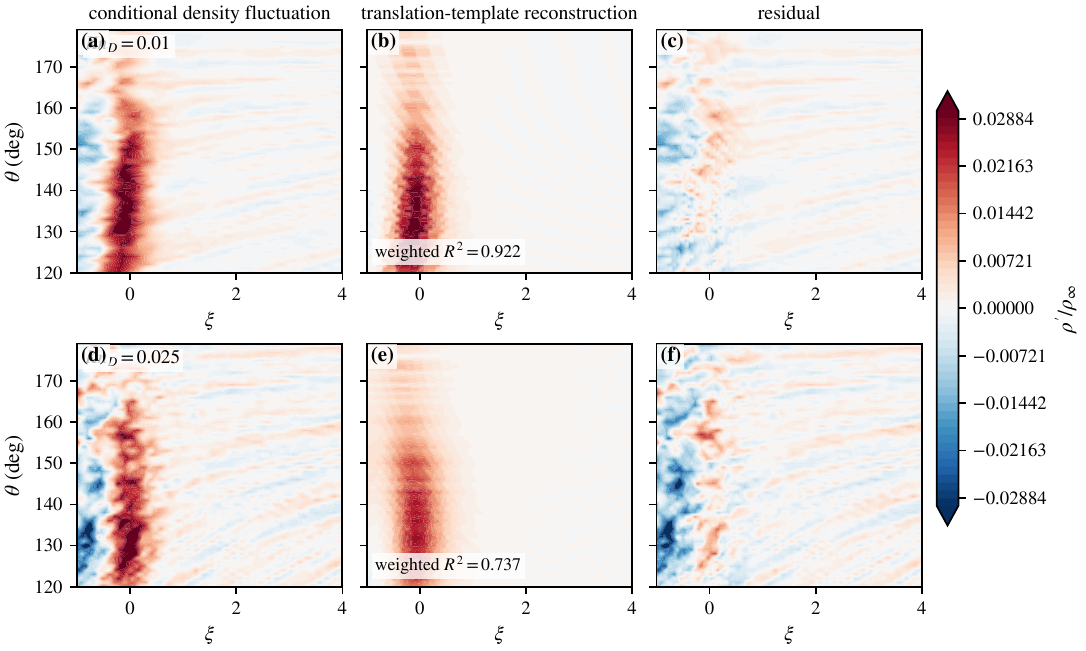}
\caption{Conditional density displacement reconstructed from the full field. The left column is half the difference between composites conditioned on the upper and lower 15\% of marker amplitude, the middle column is the fitted translation template, and the right column is the residual. The two rows correspond to $\KnD=0.01$ and $0.025$.}
\label{fig:reconstruction}
\end{figure}

Conditional composites improve the signal-to-noise ratio without changing the template. \Cref{fig:reconstruction} subtracts the mean of the most negative marker events from the mean of the most positive events. At both resolved Knudsen numbers, the actual density composite has a same-signed body-normal-ray dipole over a broad angular sector. The translation reconstruction captures the global front structure, with weighted coefficients of determination $R^2=0.922$ and 0.737, respectively; here $R^2$ is the standard variance-explained coefficient evaluated with the same spatial weights used in the matched filter. The residual retains fine angular and downstream variation but no comparable sector-wide density ridge.

\subsection{Consistency under early mean-layer broadening}
\begin{figure}[t]
\centering
\includegraphics[width=\textwidth]{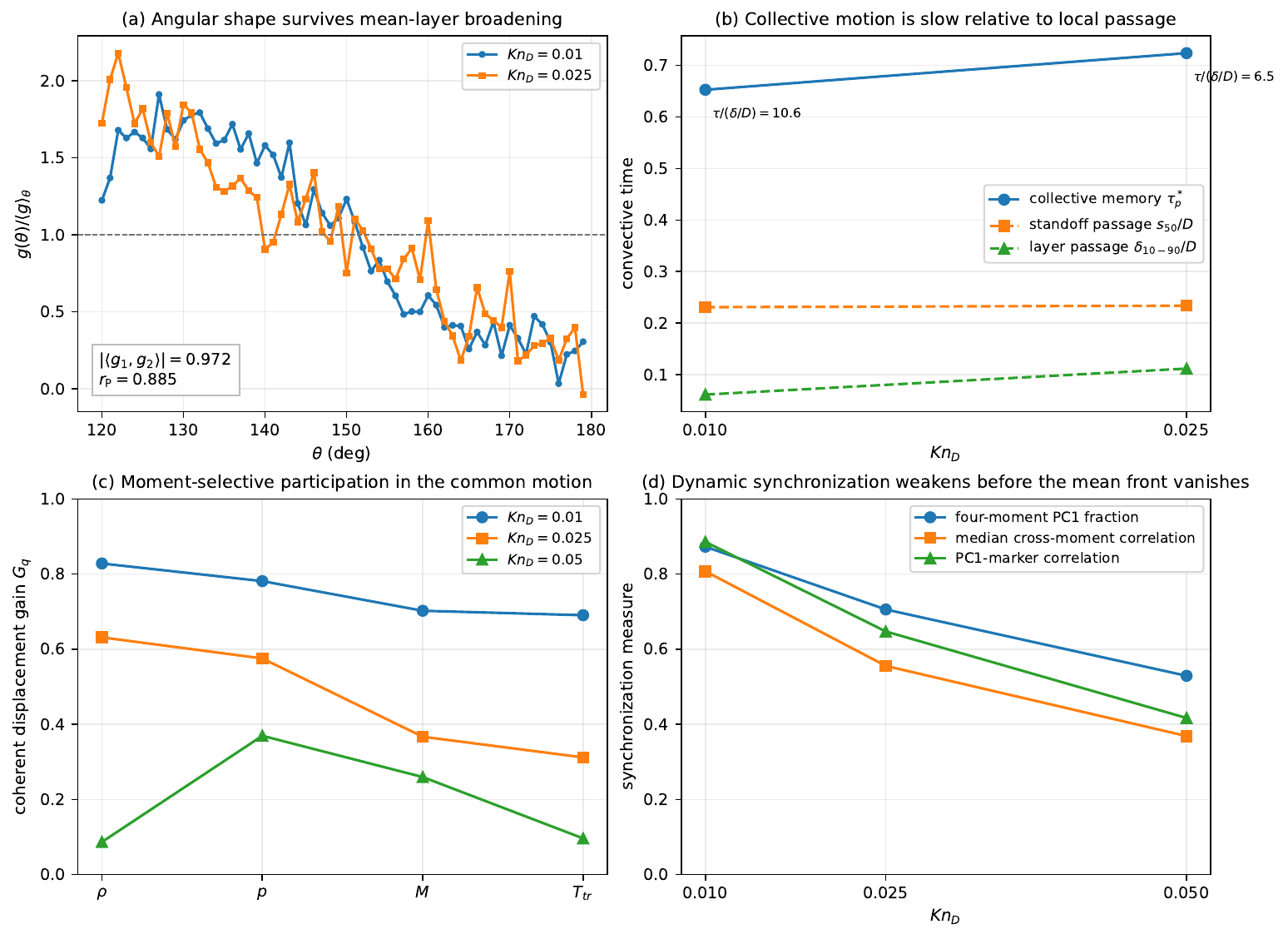}
\caption{Static broadening and dynamic organization. (a) Angular eigenfunctions at $\KnD=0.01$ and $0.025$, normalized by their sector means. Their absolute normalized inner product is $0.972$. (b) Collective memory compared with freestream-normalized mean-standoff and 10--90-width time units; these are not local residence times. (c) Coherent displacement gain $G_q$ of density, pressure, Mach number and translational temperature. (d) Multi-moment synchronization measures; the plotted label PC1 denotes the first principal component. The mean layer broadens strongly while the angular displacement geometry and slow memory persist. Reduced non-density participation is evidence consistent with moment-selective weakening, subject to observable-dependent signal-to-noise. The $\KnD=0.05$ values are diagnostic matched-filter outputs rather than evidence of a resolved collective covariance.}
\label{fig:dynamic}
\end{figure}

The principal new separation between mean and dynamic behaviour is shown in \cref{fig:dynamic}. Between $\KnD=0.01$ and $0.025$, the sector-median 10--90 density width increases from $0.123R$ to $0.224R$, an $82\%$ broadening, whereas the mean half-jump standoff changes by only $1.3\%$. The layer therefore undergoes substantial internal spreading before its centre moves appreciably. In spite of this static change, the inferred angular displacement modes remain almost collinear: their absolute normalized inner product is $0.972$ (the demeaned Pearson shape correlation is $0.885$). Both modes remain same-signed over the complete $120^\circ$--$179^\circ$ sector, although the angular envelope is not exactly uniform. Early rarefaction thus changes the thickness of the compression layer much more strongly than it changes the geometry of the resolved collective motion.

The time-scale comparison in \cref{fig:dynamic}(b) is descriptive. The inferred memories are $\tau_p^*=0.653$ and $0.724$, while the corresponding freestream-normalized width units are $t_\delta^*=0.0615$ and $0.112$. Thus the covariance memory is 10.6 and 6.5 times these nominal geometric units, and 2.83 and 3.09 times the corresponding standoff units. These ratios show that the persistent coordinate is slow relative to freestream normalization, but they are not local residence-time ratios because the normal velocity decreases through the layer. They therefore do not, by themselves, rule out an advective contribution.

\begin{table}
\centering
\caption{Dynamic consistency and multi-moment synchronization. The shape similarity is the absolute normalized inner product with the $\KnD=0.01$ mode. The ratios involving $t_\delta^*$ use freestream-normalized geometric width units, not local residence times. $S_1$ is defined in \cref{eq:synch}, and $r_{q_iq_j}$ denotes the Pearson correlation between the standardized full-field displacement amplitudes of two distinct moments $q_i$ and $q_j$.}
\label{tab:dynamic}
\small
\begin{tabular}{ccccccc}
\toprule
$\KnD$ & $\deltaten/R$ & $\tau_p^*$ & $\tau_p^*/t_\delta^*$ & shape similarity & $S_1$ & median $r_{q_iq_j}$\\
\midrule
0.010 & 0.123 & 0.653 & 10.61 & 1.000 & 0.873 & 0.807\\
0.025 & 0.224 & 0.724 & 6.46 & 0.972 & 0.706 & 0.555\\
0.050 & 0.359 & -- & -- & -- & 0.529 & 0.369\\
\bottomrule
\end{tabular}
\end{table}

In \cref{fig:dynamic}, the label PC1 denotes the first principal component of the standardized four-moment amplitude matrix introduced in \cref{eq:synch}.

\subsection{Moment-selective weakening of synchrony}
Persistence of the marker shape does not imply that every macroscopic moment continues to translate with equal amplitude. Equation~\eqref{eq:gain} separates this issue. At $\KnD=0.01$, the coherent gains are $G_\rho=0.828$, $G_p=0.781$, $G_M=0.702$ and $G_{T_{tr}}=0.690$. The four moments therefore participate in a nearly common displacement, with density and pressure slightly closer to the marker than Mach number and translational temperature. The first standardized multi-moment component contains $87.3\%$ of amplitude variance, its correlation with the marker is $0.886$, and the median pairwise moment correlation is $0.807$.

At $\KnD=0.025$, the displacement coordinate remains resolved but the participation hierarchy becomes much stronger. The density and pressure gains decrease to $0.631$ and $0.575$, while the Mach-number and translational-temperature gains fall to $0.367$ and $0.312$. The first multi-moment component now contains $70.6\%$ of the variance and the median cross-moment correlation decreases to $0.555$. This two-state contrast is evidence consistent with a collective geometry whose thermodynamic response becomes less synchronous. Because the observables have different sampling variances and attenuation bias, the gain changes are not interpreted as a universal rarefaction law. The behaviour is distinct from the static variable-dependent scale hierarchy documented in the companion paper: here the separation is measured in temporal amplitudes of a common displacement template, not in parameter-dependent mean profiles.

\begin{figure}[t]
\centering
\includegraphics[width=\textwidth]{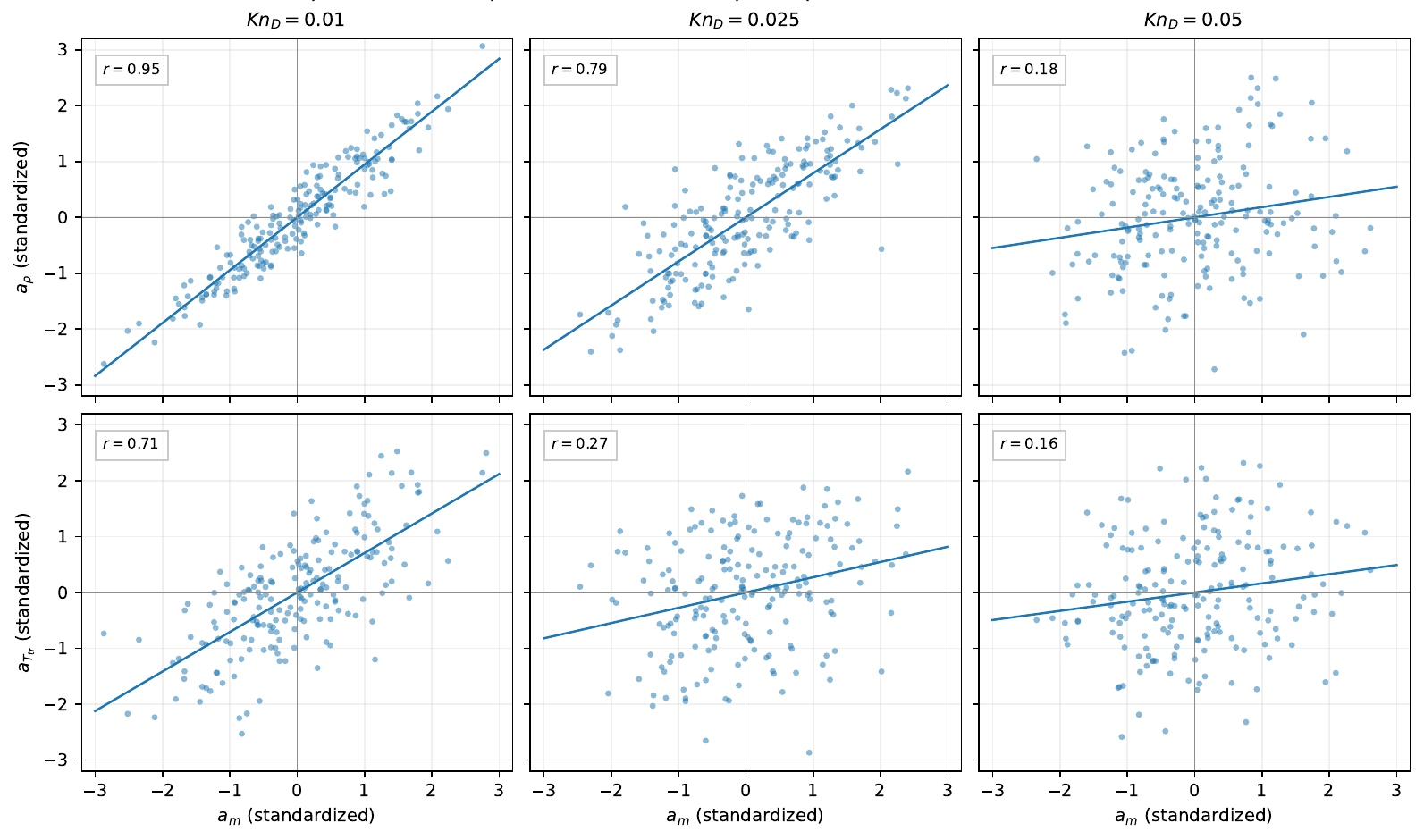}
\caption{Standardized phase portraits of marker displacement $a_m$ against density and translational-temperature equivalent displacement amplitudes. The density relation remains narrow at $\KnD=0.01$ and $0.025$, while the thermal relation loses coherence more rapidly. The $\KnD=0.05$ panels are diagnostic matched-filter outputs; the covariance mode is not classified as resolved there.}
\label{fig:phase}
\end{figure}

The phase portraits in \cref{fig:phase} make this distinction visible without energy-ranking the fields. Density remains closely slaved to the marker at $\KnD=0.01$ and $0.025$, whereas the translational-temperature cloud broadens and rotates toward weak correlation at $0.025$. At $\KnD=0.05$, neither phase portrait provides a robust common-coordinate signature under the full inference criteria. The conservative conclusion is therefore evidence for \emph{moment-selective weakening of synchrony}: early rarefaction preserves a slow compression-coordinate geometry while velocity--sound-speed and thermal amplitudes contain a larger unsynchronized component.

\subsection{Cadence-limited identifiability at $\KnD=0.05$}
The failure to resolve a persistent coordinate at $\KnD=0.05$ is also consistent with the available temporal sampling. The fitted nominal memory is $\tau_p^*=0.280$, whereas the output spacing is $\Delta t^*=0.431$, giving $\Delta t^*/\tau_p^*=1.54$ and a one-step coefficient $\phi_p=0.214$. By comparison, the resolved $\KnD=0.01$ and $0.025$ records have ratios 0.44 and 0.33 and coefficients 0.643 and 0.718. Thus the $0.05$ record samples a putative decay only after more than one relaxation time, leaving little separation from a nearly white component. This quantitative cadence limit explains why a moderate pressure--marker correlation can coexist with failed design-scale covariance identification, and it prevents the $0.05$ result from being interpreted as evidence of physical disappearance.

\subsection{What can be concluded beyond $\KnD=0.025$}
\begin{figure}[t]
\centering
\includegraphics[width=\textwidth]{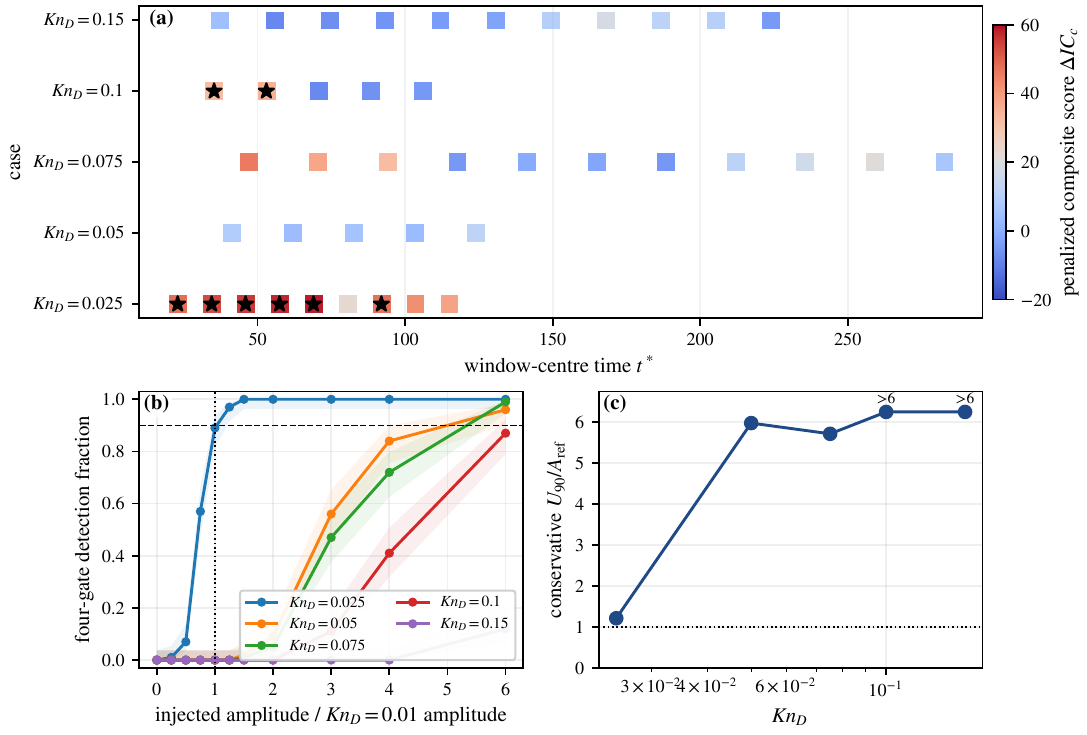}
\caption{Persistence and detectability limits. (a) Sliding-window penalized composite-score difference $\Delta IC_c$; black stars denote windows passing the combined score, leave-one-group-size-out cross-validation (LOgSO-CV), shape, far-angle and positive-semidefinite-projection criteria. (b) Four-gate detection fraction after injecting the $\KnD=0.01$ reference mode into the measured sampling covariance of each case. (c) Conservative amplitude required for the lower Wilson bound to reach 90\% detection relative to the reference amplitude. This is a sensitivity calibration for the specified injected mode, not the power of the complete final classifier.}
\label{fig:limits}
\end{figure}

A non-detection is meaningful only when the analysis could have detected a mode of relevant size. Sliding-window and injection tests are summarized in \cref{fig:limits}. The $\KnD=0.025$ mode passes the strict criterion in six of nine overlapping windows. These windows share 75\% of their samples and are not independent trials. At $\KnD=0.1$, only two early windows pass, whereas the complete record is ambiguous. No windows pass at $\KnD=0.05$, $0.075$ or $0.15$ under the combined criterion.

The injection calibration uses the $\KnD=0.01$ shape and time scale with the measured sampling covariance of each case. It tests a four-gate subset of the final classifier and therefore does not measure power against shape or model misspecification. At $\KnD=0.025$, the detection fraction at the reference amplitude $A_{ref}$ is 0.89 with a 95\% Wilson interval of 0.814--0.937. The point-estimate interpolation gives 1.03, but the conservative lower-Wilson definition used here gives $U_{90}/A_{ref}=1.21$. For $\KnD=0.05$ and $0.075$, the conservative ratios rise to 5.98 and 5.72; at $\KnD=0.1$ and $0.15$, 90\% sensitivity is not reached even at six times the reference amplitude. The noise-only experiment produced 0 detections in 100 replicates, whose 95\% Wilson upper bound is 3.7\%, rather than proving a zero false-positive probability. Consequently, the records over $0.05\leq\KnD\leq0.15$ cannot exclude persistence of a low-Knudsen-sized mode. The result is resolution at two near-continuum states, not demonstrated disappearance with rarefaction.

\section{Discussion}
\subsection{Relation to stochastic shock-motion models}
The analysis identifies two apparently contradictory facts. The instantaneous shock-attached fields are high rank, yet the repeat-run controls support a robust angular displacement. There is no contradiction because energetic dominance and dynamic organization are different properties. The translation template accounts for only $1$--$5\%$ of raw instantaneous variance, but its amplitude is correlated across distant angles, across two Knudsen-number records and across multiple macroscopic moments. The coordinate is therefore best understood as a weak slow direction embedded in a much larger broadband kinetic state space.

A useful reduced description follows the stochastic shock-motion lineage of \citet{Plotkin1975} and its later low-order developments \citep{PoggieSmits2005,Piponniau2009,TouberSandham2011,ClemensNarayanaswamy2014}. Let $a(\tstar)$ be the scalar front-displacement amplitude, $f(\tstar)$ its effective broadband forcing per unit dimensionless time, $\kappa^*$ the dimensionless restoring rate, and $\tau_a^*=1/\kappa^*$ the associated relaxation time. The leading model is
\begin{equation}
 \frac{\mathrm d a}{\mathrm d\tstar}+\kappa^*a=f(\tstar),
 \qquad \kappa^*=\frac{1}{\tau_a^*}.
 \label{eq:plotkin}
\end{equation}
For white forcing the stationary process is the Ornstein--Uhlenbeck process \citep{OrnsteinUhlenbeck1930}. Let $\omega^*$ denote dimensionless angular frequency, $S_{aa}(\omega^*)$ the displacement power spectral density and $S_{ff}$ the constant white-forcing spectral level. The Lorentzian response is
\begin{equation}
 S_{aa}(\omega^*)=\frac{S_{ff}}{(\kappa^*)^2+(\omega^*)^2}.
 \label{eq:lorentzian}
\end{equation} The present autoregressive fit is the discrete-time counterpart of \cref{eq:plotkin}. It estimates $\kappa^*$ but does not yet identify $S_{ff}$ or prove that the forcing is upstream mass flux rather than another global numerical or physical input. The analogy to Plotkin dynamics is therefore mechanistic at the level of a damped low-pass response, not a claim that a rarefied bow shock and a separated turbulent interaction share the same forcing mechanism.

\subsection{Frequency-domain and forcing-proxy stress tests}
\begin{figure}[t]
\centering
\includegraphics[width=\textwidth]{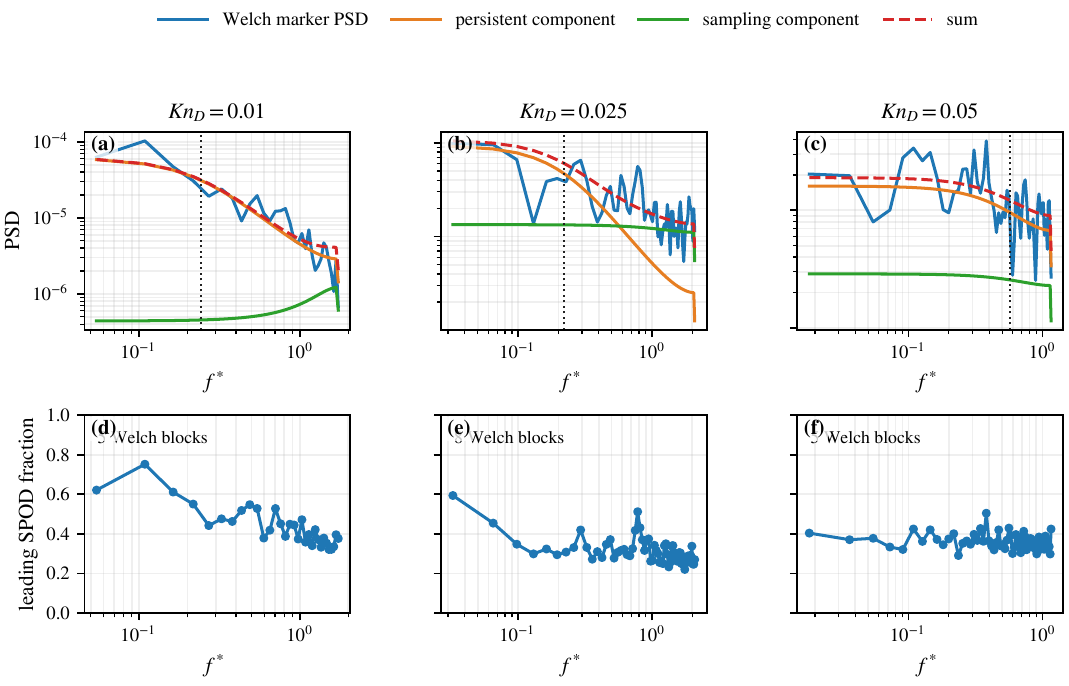}
\caption{Frequency-domain cross-check.  The upper row compares the Welch power spectral density (PSD) of the collective marker (blue) with the persistent first-order autoregressive [AR(1)] contribution, the correlated sampling contribution and their sum inferred independently from temporal coarse graining.  The dotted line marks $1/(2\pi\tau_p^*)$ when a persistent component is resolved.  The lower row shows the leading marker-field spectral proper orthogonal decomposition (SPOD) fraction.  The small number of Welch blocks is stated in each panel; the figure is used to test low-pass consistency rather than to claim a discrete frequency.}
\label{fig:freqcheck}
\end{figure}

The time-domain inference was cross-checked without refitting its parameters in the frequency domain. In this subsection, power spectral density (PSD) refers to spectral power per unit dimensionless frequency; it is unrelated to the positive-semidefinite covariance projection used in the statistical fit.  For a discrete first-order autoregressive [AR(1)] component, let $\phi$ be the lag-one coefficient, $Q_{\mathrm{AR}}$ the scalar innovation variance, $f^*$ the dimensionless cyclic frequency, $\Delta t^*$ the sampled dimensionless time step and $\mathrm{i}=\sqrt{-1}$. Its exact sampled power spectral density is proportional to
\begin{equation}
 S_{\mathrm{AR1}}(f^*)=\frac{Q_{\mathrm{AR}}}{\left|1-\phi\exp(-2\pi\mathrm{i} f^*\Delta t^*)\right|^2}.
 \label{eq:ar1spectrum}
\end{equation} The persistent and sampling spectra predicted from the coarse-graining fit are added and compared with a Welch estimate of the measured collective marker in \cref{fig:freqcheck}.  The agreement is broadband rather than resonant: the root-mean-square logarithmic discrepancy is approximately $0.21$ decades in the three displayed records.  The two resolved cases show enhanced low-frequency content compatible with a low-pass response, but no narrow spectral line.  A marker-field spectral proper orthogonal decomposition (SPOD) cross-check likewise concentrates its largest leading fraction toward the low-frequency end, but only five Welch blocks are available at $\KnD=0.01$ and $0.05$ and eight at $0.025$.  SPOD is therefore supporting evidence for broadband organization, not an independent claim of a selected oscillatory mode.

The attached-field caches also permit exploratory upstream proxies to be formed in the region $\xi\geq3.3$: density, pressure and $\rho M\sqrt{\gamma R_sT_{tr}}$ as a mass-flux-magnitude proxy. Because this region is defined in the moving attached coordinate, however, it follows the inferred front and does not sample a fixed laboratory-frame plane; shock displacement can therefore contaminate the proxy by changing which fluid parcel is observed.  Across $\KnD=0.01$, $0.025$ and $0.05$, neither lagged correlation nor low-frequency coherence produces a circular-shift-corrected significant and repeatable forcing signature.  The existing data therefore identify a damped low-pass response but not the source of $f(t^*)$.  In particular, the manuscript does not attribute the forcing to freestream mass-flux fluctuations.  A dedicated upstream sampling plane is required for that mechanism test.

The time-scale comparison is deliberately limited. At the two resolved conditions the memory is $6.5$--$10.6$ times the freestream-normalized mean-width unit and about three times the analogous standoff unit. These are not local residence times, because the gas decelerates strongly through the bow layer; therefore the ratios establish a slow nominal memory but do not rule out advection. Let $\mathcal L_{\mathrm{kin}}$ denote a stable linearized kinetic evolution operator and $\eta(t)$ stochastic forcing acting on the fluctuation state $q'$. A general representation is
\begin{equation}
 \frac{\partial q'}{\partial t}=\mathcal L_{\mathrm{kin}} q'+\eta(t).
 \label{eq:stochastic_operator}
\end{equation}
Let $\bm C$ denote the stationary state covariance, $\bm Q_\eta$ the covariance of $\eta$, and $\dagger$ the adjoint. An ideal linear system then satisfies
\begin{equation}
 \mathcal L_{\mathrm{kin}}\bm C+\bm C\mathcal L_{\mathrm{kin}}^\dagger+\bm Q_\eta=0.
 \label{eq:lyap}
\end{equation} The present work estimates a projected macroscopic part of $\bm C$; it does not identify $\mathcal L_{\mathrm{kin}}$ or $\bm Q_\eta$ separately.

\subsection{Dynamic consistency is distinct from mean similarity}
The companion paper \citep{RoohiShoja2026Mean} asks whether the \emph{mean} rarefied bow layer can be represented by one shifted and broadened profile across parameter space. It reports mean inflation, variable-specific length scales and breakdown of universal static similarity. The present result is complementary and non-redundant. Between $\KnD=0.01$ and $0.025$ the mean density width increases by $82\%$, yet the normalized displacement eigenfunction and convective memory remain statistically compatible. The dynamic coordinate is therefore not a trivial restatement of the mean profile collapse. It describes how the instantaneous layer fluctuates about each mean state.

This separation suggests a two-level reduced description. The first level is a slowly varying mean geometry parameterized by rarefaction, as quantified in the companion study. The second is a stochastic displacement amplitude that acts on the local mean gradients. Let $a(t;\KnD)$ be the scalar collective displacement amplitude at a given Knudsen number, $g(\theta;\KnD)$ its normalized angular shape, $\partial_s\mean q$ the wall-normal gradient of the local mean field, and $q'_\perp$ the high-rank residual outside the retained displacement coordinate. Such a model has the schematic form
\begin{equation}
 q(s,\theta,t;\KnD)
 =\mean q(s,\theta;\KnD)
 -a(t;\KnD)g(\theta;\KnD)\partial_s\mean q
 +q'_\perp.
 \label{eq:twolevel}
\end{equation} The present data show that $g$ changes much less rapidly than $\mean q$ between the two resolved Knudsen numbers. This is precisely why a weak coordinate can remain dynamically recognizable even as the mean layer broadens substantially.

\subsection{Moment-selective displacement and relaxation}
The coherent gains reveal a hierarchy that is hidden by marker-only analysis. Density is the variable used to define the half-jump centre, so a strong density gain is expected, but pressure follows almost as closely. This is physically consistent with a compression displacement in which density and translational energy jointly alter pressure. Mach number is more weakly coupled because it depends on both velocity deceleration and the local sound speed. Translational temperature is weakest at $\KnD=0.025$, consistent with thermal adjustment being distributed over a broader relaxation region than the density compression.

At $\KnD=0.05$, pressure retains a larger marker correlation ($0.510$) than density ($0.183$), but this does not rescue a displacement interpretation.  The unshifted pressure template explains only $0.77\%$ of weighted instantaneous variance and is not the best spatial shift; a template displaced by $\Delta\xi=1$ performs slightly better.  Density behaves even less locally, with its best shifted template explaining more than three times the zero-shift projection.  Together with a two-component LOgSO-CV ratio above unity and a $54\%$ positive-semidefinite projection correction, the pressure correlation is interpreted as diffuse thermodynamic co-variation rather than translation of the mean compression gradient.

The key point is not simply that the correlations decrease. The angular displacement shapes are strongly aligned across the two resolved cases while the relative moment amplitudes separate. We describe this two-state contrast as evidence consistent with moment-selective weakening of synchrony. It is measured from synchronized fluctuation amplitudes rather than mean thicknesses, but differences in observable-specific sampling variance and regression attenuation can contribute to the gain change. The data therefore support the hypothesis that the compression geometry remains organized while thermal and velocity--sound-speed responses contain a larger unsynchronized component at $\KnD=0.025$; they do not yet establish a universal rarefaction trend.

\subsection{Relation to kinetic stability theory}
The kinetic linear stability theory (kLST) framework of \citet{Karpuzcu2026} linearizes the Boltzmann equation with a Bhatnagar--Gross--Krook (BGK) collision model about one-dimensional normal shocks and demonstrates that non-Maxwellian velocity distributions shift the eigenspectrum toward less stable regions. Their calculation concerns planar argon shocks at $M_\infty\leq4$ and deterministic spanwise perturbations, whereas the present calculation concerns stochastic, curved, body-coupled nitrogen flow at $M_\infty=10$. Direct equality between their eigenmodes and the present covariance mode would therefore be unjustified.

The two studies nevertheless define a useful next step. Along the stagnation line, one may construct a locally planar kinetic base state and determine whether its least-damped response has a macroscopic projection proportional to $-\partial_s\mean q$. Agreement in spatial shape and decay time would support the interpretation of the measured coordinate as stochastic excitation of a stable kinetic response. Disagreement would be equally informative, indicating that curvature, wall coupling or non-normal forcing is essential. The present displacement template supplies the observable against which such a theory should be tested.

\subsection{Implications for particle-data modal analysis}
The physical-support audit provides a general warning for registered particle fields. A map can be numerically smooth and produce a large singular value while representing repeated extrapolated values rather than fluid motion. Support must be enforced before interpolation, and modal energy must be audited spatially before it is interpreted. This issue is especially acute near curved walls, moving fronts and adaptive point clouds.

A second warning is that finite-particle noise is not white. Snapshot accumulation, particle persistence and adaptive sampling produce correlated feature error. Group averaging after feature extraction would not reproduce the statistics of a nonlinear marker; the fields must be averaged first and the marker re-extracted. The attenuation function in \cref{eq:attenuation} is then a first-order covariance model, not an exact identity for the re-extracted crossing. Its limitations must be exposed through calibration sensitivity and independent seed/loading repeats. The combined use of coarse graining, design-scale prediction, synthetic controls and full-field matched filtering is more informative than applying proper orthogonal decomposition (POD), dynamic mode decomposition (DMD) or spectral proper orthogonal decomposition (SPOD) alone.

For reduced modelling, \cref{eq:twolevel} suggests that one should not seek a globally low-rank representation of every instantaneous field. A more efficient strategy is to represent the parameterized mean layer, the slow displacement amplitude and the broadband residual separately. Density and pressure may share a displacement coordinate over the near-continuum range, whereas Mach number and thermal variables require additional latent coordinates or relaxation states.

\subsection{Limitations and scope of the conclusions}
The study uses one body shape, Mach number, wall condition and rotational nitrogen model. The perfect-gas temperature estimate above shows that vibration is energetically accessible, so vibrational relaxation could add a slow coordinate in physical nitrogen. The present result must therefore not be generalized to thermochemical nonequilibrium without a vibrationally active calculation.

Absolute fluctuation amplitudes depend on simulator-particle weight, sampling duration and output cadence; they are not universal molecular fluctuation levels. The completed $2\times2$ campaign with independent seeds and particle levels $N_p$ and $2N_p$ reproduces the normalized response shape (pairwise alignments 0.967--0.990) and gives overlapping block-resampling memory ranges. At the same time, doubling the loading reduces the raw marker variance to 0.457--0.479 and the inferred persistent variance to 0.491--0.517 of their $N_p$ values. The controls therefore support robustness of classification, geometry and memory to realization and loading, but not invariance of absolute amplitude. They do not convert the measured displacement variance into an absolute fluctuation amplitude of real nitrogen.

The present analysis uses only the upper upstream sector. It therefore addresses the symmetric-sector raywise displacement subspace and cannot detect a full-domain antisymmetric rocking component. Distinguishing collective expansion from spontaneous rocking requires a full-domain analysis. The full-field template is also a first-order translation model and explains only a small fraction of unconditioned variance; curvature changes, width fluctuations and moment-specific relaxation can contribute additional coordinates. Finally, the strong alignment of $|g|$ with inverse density-gradient magnitude means that confirmation with a distinct gradient-, pressure- or Mach-based marker remains an important robustness test.

Finally, the mode is resolved only at $\KnD=0.01$ and $0.025$. The calibrated sensitivity analysis over $0.05\leq\KnD\leq0.15$ cannot exclude a mode of comparable low-Knudsen amplitude, so the manuscript does not claim disappearance or a critical Knudsen number. The $\KnD=0.25$, $0.5$ and 1 records enter the common-200 field-POD audit but not the full detectability calibration. Direct velocity-distribution-function analysis or kinetic resolvent/eigenmode calculations would be needed to connect the measured macroscopic coordinate rigorously to a collision-mediated mechanism.

\section{Conclusions}
Time-resolved direct simulation Monte Carlo (DSMC) fields of Mach-10 rotationally relaxing nitrogen flow over a circular cylinder were analysed for evidence of a persistent bow-layer displacement distinct from correlated particle-sampling fluctuations. The principal conclusions are as follows.
\begin{enumerate}[leftmargin=1.6em]
\item Shock-attached maps require physical support. Solid-side filling by nearest neighbours produced false leading proper orthogonal decomposition (POD) energies of order $40$--$50\%$. After gas-domain clipping, the leading combined-state energy is only $1.1$--$4.3\%$ and 146--172 modes are required for $90\%$ variance.
\item A same-signed persistent angular component is resolved at $\KnD=0.01$ and $0.025$. It has positive far-angle covariance and memories $\tau_p^*=0.653$ and $0.724$. In the completed $2\times2$ audit, all four seed/loading records lie above their case-matched 99th-percentile null thresholds, the pairwise angular-mode alignments are 0.967--0.990 and the memory ranges overlap. Doubling the loading nevertheless reduces both raw and inferred persistent variance by approximately one half, so the supported robustness concerns classification, normalized shape and memory rather than absolute amplitude.
\item The mean 10--90 density width increases by $82\%$ between the two resolved states, while the angular modes retain an absolute normalized inner product of $0.972$. The memory is $6.5$--$10.6$ times a freestream-normalized geometric width unit and about three times the analogous standoff unit. These are not local residence-time ratios and do not by themselves exclude advection.
\item Density, pressure, Mach number and translational temperature provide complementary full-field checks. They share simulator particles and are not independent experiments. Density and pressure have the strongest coherent displacement gains; the faster reduction of Mach-number and translational-temperature participation is evidence consistent with moment-selective weakening, subject to observable-dependent signal-to-noise and attenuation.
\item The collective coordinate is weak, explaining only $1$--$5\%$ of raw instantaneous field variance. It is a slow direction embedded in broadband kinetic fluctuations and is not evidence of a newly discovered discrete oscillation or a linear instability.
\item The principal methodological contribution is the combined audit of physical interpolation support, correlated sampling covariance, angular collective structure, complementary full-field response, completed seed/loading repeats and identifiability limits. The novelty is not the first observation of low-frequency shock fluctuation, and the $82\%$ broadening is context supplied by the companion mean-flow study.
\item Over $0.05\leq\KnD\leq0.15$, calibrated sensitivity deteriorates too rapidly to demonstrate disappearance. The higher $\KnD=0.25$, 0.5 and 1 cases enter the common-200 POD audit but not the full exclusion calibration. No critical Knudsen number is claimed.
\item The small amplitude and strong correlation of the angular envelope with inverse density-gradient magnitude leave marker sensitivity as a residual limitation. Verification with a distinct pressure-, Mach- or gradient-based marker and a full-domain calculation would further test the interpretation and permit antisymmetric rocking to be assessed.
\end{enumerate}

\section*{Funding}
This research received no specific grant from any funding agency, commercial or not-for-profit sectors.

\section*{Declaration of interests}
The authors report no conflict of interest.

\section*{Author contributions}
Ahmad Shoja-Sani performed the DS2V production simulations, curated the time-resolved data, implemented campaign-level post-processing and contributed to the physical analysis and first manuscript draft. Ehsan Roohi conceived and supervised the study, developed the noise-separation and full-field interpretation, and led manuscript revision. Both authors reviewed and approved the manuscript. Ehsan Roohi is the corresponding author.

\section*{Prior publication and data provenance}
The authors' Physics of Fluids article \citep{RoohiShojaAzghadi2026PoF} previously reported neural-operator surrogates for steady cylinder fields, including a fixed-Mach-10 cross-Knudsen argon study and a separate nitrogen Mach-number study. The present paper does not republish those surrogate predictions or performance results. Its new contribution is the time ordering and stochastic analysis of consecutive nitrogen DSMC output blocks, including physical-support auditing, correlated-noise separation, relaxation memory, power limits and multi-moment displacement validation.

\section*{Data availability}
The manuscript source, processed summary tables and figure-generation code are available in the \href{https://github.com/Ehsan-Roohi/DSMC_TimeResolved_BowShock}{project repository}. The present submission package contains the complete Overleaf build, the processed CSV files used by the figures, and the processed $2\times2$ audit tables, matched synthetic controls, DS2V input cards, seed records, random-number-generator states and executable hash. Multi-gigabyte raw DS2V snapshots, the frozen covariance-inference environment and the complete independent-seed/particle-loading output fields are available from the corresponding author because they are too large for the submission archive. The packaged audit supports the numerical claims in \cref{tab:controlcases,tab:controlpairs}; it is not an end-to-end rerunnable DSMC campaign.

\appendix
\renewcommand{\theHsection}{appendix.\Alph{section}}
\section{Support and registration diagnostics}
The support audit in \cref{fig:support} was repeated for density, Mach number, pressure and both temperature fields. The strongest false condensation occurred in the density field at $\KnD=0.25$ and the rotational-temperature field at $\KnD=0.5$--1. Across the full audit, 88--99.8\% of the apparent leading-mode energy in the most affected cases lay in the non-physical attached support. Corrected results were insensitive to physical buffers $s/R=0$, 0.02 and 0.05 within the retained gas region. The common-200 comparisons use $s/R\geq0.02$.

\section{Statistical decision rules}
The covariance representation is not accepted from $\Delta IC_c$ alone. A resolved classification requires: (i) positive block-resampled composite-score preference; (ii) lower leave-one-group-size-out error for the persistent-plus-sampling representation; (iii) positive far-angle covariance; (iv) a smooth displacement-like mode; (v) controlled false detection under sampling-only synthetic data; (vi) a small correction in the positive-semidefinite-constrained covariance fit; and (vii) repeat-run seed/loading robustness. This combined rule is why the nominal positive $\Delta IC_c$ at $\KnD=0.05$ is rejected.

The corrected injection analysis uses a four-gate criterion rather than raw negative eigenvalue mass of the unconstrained covariance. The sampling-only experiment yields 0 detections in 100 replicates; the corresponding 95\% Wilson upper bound is 3.7\%, so the underlying false-positive probability is not asserted to be zero. At the reference amplitude the $\KnD=0.025$ detection fraction is 0.89 [0.814, 0.937]. Because the injection uses the $\KnD=0.01$ reference shape and does not reproduce every final gate, it is a specified-mode sensitivity calibration rather than the power of the complete classifier.

\subsection{Calibrated $2\times2$ repeat audit}
The completed control campaign contains four 600-block records. Cases N1-A and N1-B use $N_p=1.5\times10^6$ simulator particles; N2-A and N2-B use $2N_p=3.0\times10^6$. The suffixes A and B denote seeds 104729 and 130363, respectively. For every case, $\Delta IC_c$ is compared directly with 400 case-matched sampling-only synthetic records. The archived audit tables retain the historical column name \texttt{delta\_aicc}; throughout the manuscript the quantity is correctly denoted $\Delta IC_c$ and is not interpreted as a likelihood-based Akaike statistic.

\begin{center}
\captionof{table}{Casewise calibrated repeat-run audit. $q_{0.99}^{0}$ is the 99th percentile of the case-matched sampling-only $\Delta IC_c$ distribution, $p_{\mathrm{MC}}$ is the finite Monte Carlo exceedance probability and $R_{cv}$ is the two-component/noise-only LOgSO-CV error ratio. The final three columns are 2.5th-percentile block-resampling bounds for far-angle correlation and uniform-mode correlation, followed by the relative positive-semidefinite projection correction.}
\label{tab:controlcases}
\scriptsize
\begin{tabular}{@{}ccccccccc@{}}
\toprule
Case & loading & seed & $\Delta IC_c$ & $q_{0.99}^{0}$ & $p_{\mathrm{MC}}$ & $R_{cv}$ & $r_{far,0.025}$ & $r_{unif,0.025}$ / $\epsilon_{proj}$ \\
\midrule
N1-A & $N_p$  & 104729 & 63.82 & 20.87 & 0.00249 & 0.355 & 0.0469 & 0.775 / 0.0106 \\
N1-B & $N_p$  & 130363 & 74.44 & 23.51 & 0.00249 & 0.276 & 0.0561 & 0.784 / 0.0049 \\
N2-A & $2N_p$ & 104729 & 78.77 & 33.18 & 0.00249 & 0.318 & 0.1543 & 0.839 / 0.0126 \\
N2-B & $2N_p$ & 130363 & 98.65 & 33.70 & 0.00249 & 0.394 & 0.1721 & 0.893 / 0.0426 \\
\bottomrule
\end{tabular}
\end{center}

\begin{center}
\captionof{table}{Pairwise repeat-run audit. Mode alignment is the absolute normalized inner product of the two inferred angular modes. Each of the four pairwise comparisons has overlapping block-resampling memory ranges. Variance ratios use the second case divided by the first; $V_{raw}$ is raw marker variance and $\operatorname{tr}(C_p)$ is the inferred persistent covariance trace.}
\label{tab:controlpairs}
\scriptsize
\begin{tabular}{@{}lccc@{}}
\toprule
Comparison & mode alignment & memory-range overlap & $\tau_{p,b}^*/\tau_{p,a}^*$ \\
\midrule
N1-A vs N1-B (seed at $N_p$) & 0.990 & yes & 1.165 \\
N2-A vs N2-B (seed at $2N_p$) & 0.977 & yes & 0.722 \\
N1-A vs N2-A (loading, seed 104729) & 0.967 & yes & 1.478 \\
N1-B vs N2-B (loading, seed 130363) & 0.974 & yes & 0.916 \\
\bottomrule
\end{tabular}

\medskip
\begin{tabular}{@{}lcc@{}}
\toprule
Comparison & $V_{raw,b}/V_{raw,a}$ & $\operatorname{tr}(C_{p,b})/\operatorname{tr}(C_{p,a})$ \\
\midrule
N1-A vs N1-B (seed at $N_p$) & 1.049 & 1.103 \\
N2-A vs N2-B (seed at $2N_p$) & 1.099 & 1.160 \\
N1-A vs N2-A (loading, seed 104729) & 0.457 & 0.491 \\
N1-B vs N2-B (loading, seed 130363) & 0.479 & 0.517 \\
\bottomrule
\end{tabular}
\end{center}

The repeat results therefore separate two claims that should not be conflated. The classification, normalized angular shape and memory are robust to seed and loading under the reported uncertainty ranges. The absolute variance is not invariant to simulator-particle loading and is not interpreted as a universal molecular fluctuation level.

\section{Noise-memory, gate-threshold and forcing-proxy sensitivity}
\begin{figure}[p]
\centering
\includegraphics[width=.93\textwidth]{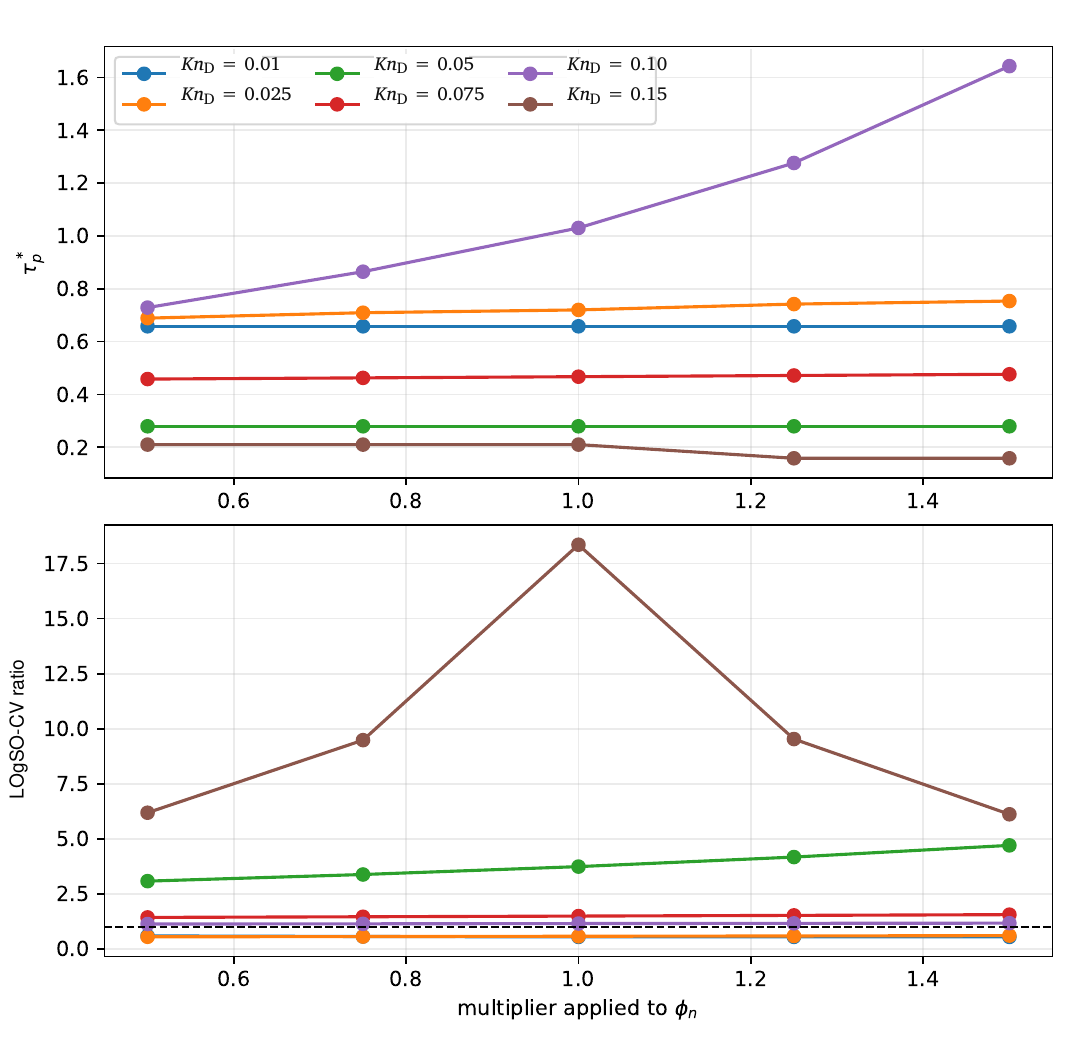}
\caption{Sensitivity to the sampling-memory calibration.  The upper panel gives the inferred physical memory and the lower panel the two-component/noise-only leave-one-group-size-out cross-validation (LOgSO-CV) error ratio as $\phi_n$ is multiplied by the stated factor.  The resolved classifications at $\KnD=0.01$ and $0.025$ survive the full tested range.}
\label{fig:phisens}
\end{figure}

The width channel is used to calibrate the sampling-memory coefficient $\phi_n$, so its transfer to the nonlinear centre marker is an identifiability assumption rather than a theorem. At $\KnD=0.01$, $\phi_n=-0.25$ is on the lower search boundary; the width variance ratios from $m=1$ to 2 are 0.186 pointwise and 0.363 for the angular mean, which cannot both be represented by one AR(1) coefficient. We therefore repeat the complete covariance inference after multiplying the calibrated value by $0.5$, $0.75$, $1$, $1.25$ and $1.5$, and after absolute shifts up to $\pm0.15$. As shown in \cref{fig:phisens}, $\KnD=0.01$ and $0.025$ satisfy the complete resolved-mode criterion for every tested perturbation. The $0.05$, $0.10$ and $0.15$ records never pass. The $0.075$ record passes only after the most extreme absolute reduction of $\phi_n$, confirming that its nominal response is calibration-sensitive rather than robust. The successful seed/loading repeats provide the separate empirical robustness check; the sensitivity grid does not make the AR(1) approximation exact.

\begin{figure}[p]
\centering
\includegraphics[width=.93\textwidth]{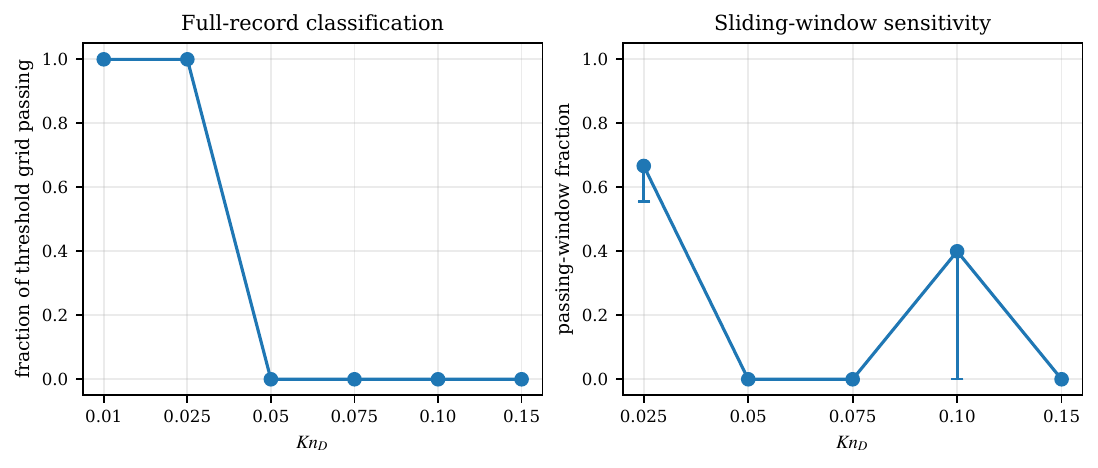}
\caption{Sensitivity to the combined detection gates.  The surfaces/curves span the stated $\Delta IC_c$, mode-alignment and positive-semidefinite-projection thresholds.  The identity of the two complete-record resolved cases is unchanged; only the fraction of passing windows in the already window-dependent records varies.}
\label{fig:gatesens}
\end{figure}

The statistical decision rule is also varied over $\Delta IC_c$ thresholds 5--15, reference-mode alignment thresholds 0.60--0.80 and positive-semidefinite-projection thresholds 0.10--0.30.  The full-record classification is invariant over this grid: only $\KnD=0.01$ and $0.025$ pass.  The sliding-window pass fraction at $0.025$ remains between 0.56 and 0.67; $0.05$, $0.075$ and $0.15$ remain zero, while $0.10$ remains restricted to early windows.  \Cref{fig:gatesens} demonstrates that the two resolved cases were not selected by tuning one threshold after examining the data.

\begin{figure}[p]
\centering
\includegraphics[width=.93\textwidth]{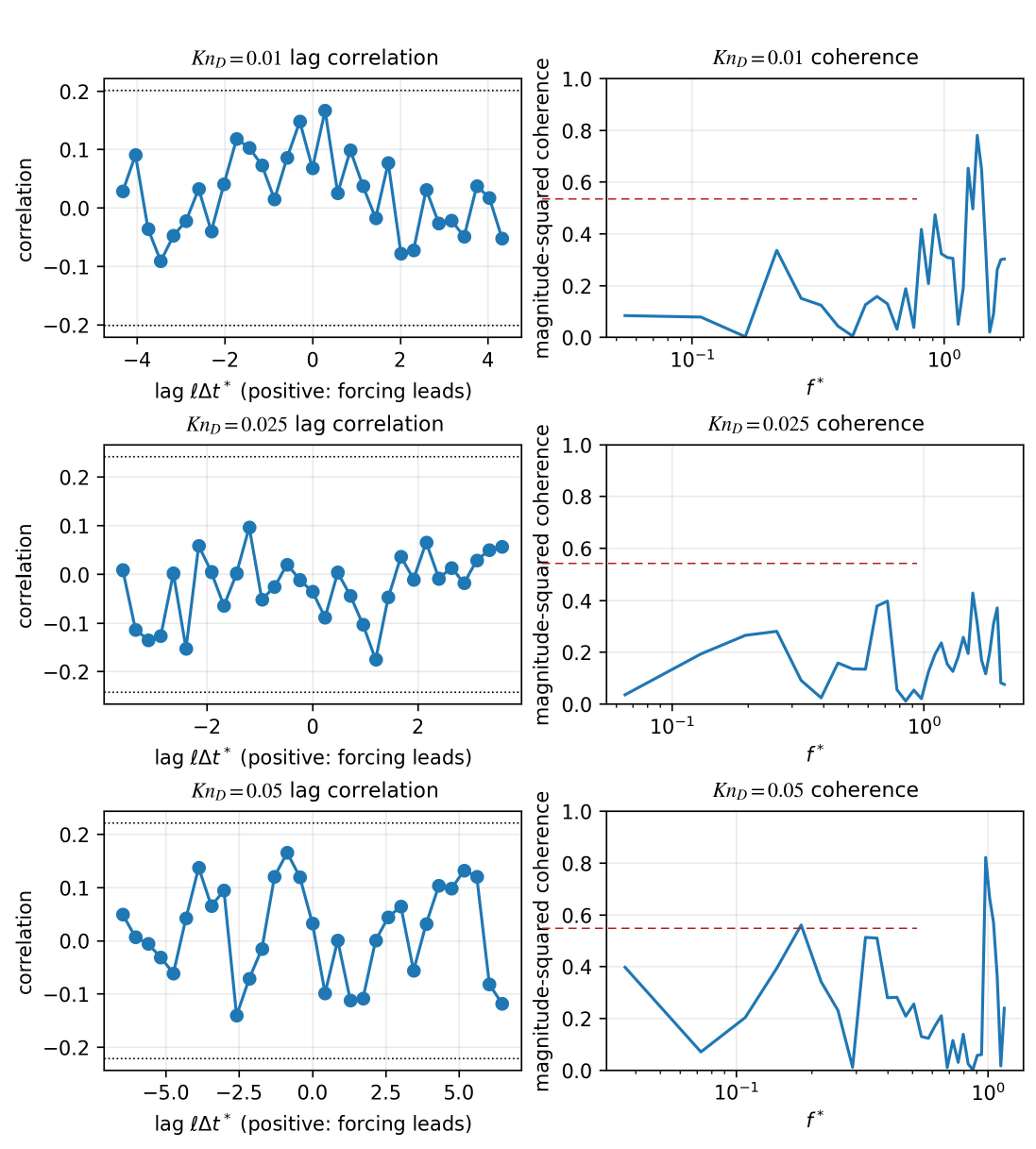}
\caption{Exploratory upstream-forcing proxies formed from the far-upstream part of the attached caches. Lagged correlations are assessed against circular-shift nulls. The horizontal dashed line in each coherence panel is the nominal 95\% threshold $1-0.05^{1/(K-1)}=0.527$ for $K=5$ Welch segments; overlap between segments makes this an optimistic guide rather than a formal independent-block test. The proxy is defined in the moving attached frame, not on a fixed laboratory plane. No density, pressure or mass-flux proxy gives a statistically significant and repeatable mechanism signature.}
\label{fig:forcingproxy}
\end{figure}

\Cref{fig:forcingproxy} reports the upstream-proxy test described in the discussion.  The largest lagged correlations are compared with circular-shift null thresholds, and the low-frequency coherence is shown only as an exploratory statistic.  No proxy produces a significant, repeatable lag and phase pattern across records.  This negative result is scientifically useful: it prevents the low-pass response from being over-interpreted as evidence for a specific upstream forcing source.

The nominal $\Delta IC_c=76.4$ at $\KnD=0.01$ lies above the 97.5th percentile (73.2) of its moving-block resampling distribution.  This is not an algebraic inconsistency: finite moving blocks omit some long-range combinations and shift the resampled statistic downward.  We therefore describe those brackets as percentile ranges of the block-resampling distribution, not as bias-corrected confidence intervals centred on the original point estimate.

\section{Detailed full-field moment diagnostics}
\Cref{tab:fullfielddetail} reports the full-field metrics from which the coherent gains are formed. The amplitude standard deviations are expressed as equivalent radial displacement divided by $R$. The projection fraction is the median squared weighted spatial correlation of the instantaneous field with the translation template; it is not a fraction of thermodynamic energy.

\begin{center}
\captionof{table}{Detailed displacement-template results. Brackets give 95\% moving-block intervals for field--marker correlation. The information is split into two blocks to avoid crowding.}
\label{tab:fullfielddetail}
\scriptsize
\begin{tabular}{@{}ccccc@{}}
\toprule
$\KnD$ & variable & $r(a_q,a_m)$ & 95\% interval & $G_q$\\
\midrule
0.010 & $\rho$   & 0.946 & [0.921,0.962] & 0.828\\
      & $p$       & 0.901 & [0.845,0.928] & 0.781\\
      & $M$       & 0.755 & [0.670,0.821] & 0.702\\
      & $T_{tr}$  & 0.707 & [0.599,0.777] & 0.690\\
\addlinespace
0.025 & $\rho$   & 0.790 & [0.723,0.847] & 0.631\\
      & $p$       & 0.727 & [0.652,0.788] & 0.575\\
      & $M$       & 0.395 & [0.253,0.512] & 0.367\\
      & $T_{tr}$  & 0.273 & [0.155,0.371] & 0.312\\
\addlinespace
0.050 & $\rho$   & 0.183 & [0.054,0.340] & 0.087\\
      & $p$       & 0.510 & [0.428,0.605] & 0.369\\
      & $M$       & 0.325 & [0.179,0.433] & 0.260\\
      & $T_{tr}$  & 0.164 & [0.034,0.284] & 0.096\\
\bottomrule
\end{tabular}

\medskip
\begin{tabular}{@{}cccc@{}}
\toprule
$\KnD$ & variable & $\sigma_{a_q}/R$ & median projection (\%)\\
\midrule
0.010 & $\rho$   & $4.82\times10^{-4}$ & 5.29\\
      & $p$       & $4.77\times10^{-4}$ & 4.03\\
      & $M$       & $5.12\times10^{-4}$ & 2.30\\
      & $T_{tr}$  & $5.37\times10^{-4}$ & 1.24\\
\addlinespace
0.025 & $\rho$   & $6.81\times10^{-4}$ & 2.37\\
      & $p$       & $6.74\times10^{-4}$ & 1.60\\
      & $M$       & $7.90\times10^{-4}$ & 1.73\\
      & $T_{tr}$  & $9.71\times10^{-4}$ & 0.97\\
\addlinespace
0.050 & $\rho$   & $6.47\times10^{-4}$ & 0.67\\
      & $p$       & $9.89\times10^{-4}$ & 0.77\\
      & $M$       & $1.09\times10^{-3}$ & 1.06\\
      & $T_{tr}$  & $8.00\times10^{-4}$ & 0.46\\
\bottomrule
\end{tabular}
\end{center}

\section{Full-record covariance inference across the transition range}
The complete-record results in \cref{tab:fullcov} explain why positive nominal composite-score preference is insufficient. The $\KnD=0.05$ and $0.075$ fits have positive $\Delta IC_c$, but their two-component LOgSO-CV ratios exceed unity and their block-resampled far-angle ranges include zero. At $0.10$, the nominal time scale has a very broad resampling upper tail. These cases are therefore not classified as resolved persistent components.

\begin{center}
\captionof{table}{Full-record correlated-noise diagnostics for the transition-range cases. The ``resampling range'' column contains percentiles of the moving-block resampling distribution. $R_{cv}$ is the two-component/noise-only LOgSO-CV error ratio and $\epsilon_{proj}$ is the relative positive-semidefinite projection correction.}
\label{tab:fullcov}
\scriptsize
\begin{tabular}{cccccccc}
\toprule
$\KnD$ & $\Delta IC_c$ & resampling range & $R_{cv}$ & $\tau_p^*$ & $r_{unif}$ & far-angle range & $\epsilon_{proj}$\\
\midrule
0.025 & 59.1 & [29.0,64.4] & 0.58 & 0.724 & 0.876 & [0.120,0.287] & $<10^{-12}$\\
0.050 & 18.0 & [1.9,37.5] & 3.74 & 0.280 & 0.767 & [-0.002,0.089] & 0.539\\
0.075 & 36.4 & [-1.8,58.5] & 1.50 & 0.466 & 0.644 & [-0.001,0.108] & 0.152\\
0.100 & 18.9 & [-16.9,51.7] & 1.16 & 1.007 & 0.584 & [-0.031,0.112] & 0.019\\
0.150 & -5.0 & [-8.2,12.9] & 18.77 & 0.210 & 0.475 & [-0.028,0.049] & 0.836\\
\bottomrule
\end{tabular}
\end{center}

\section{Derivation of the coherent-gain and memory diagnostics}
For completeness, let $G$ denote a trial scalar gain and let $\sum_t$ denote a sum over all sampled times. The least-squares estimate $\widehat G_q$ of the slope of field-equivalent displacement $a_q$ on marker amplitude $a_m$ is
\begin{equation}
 \widehat G_q=\arg\min_G\sum_t(a_q-Ga_m)^2
 =\frac{\operatorname{cov}(a_q,a_m)}{\operatorname{var}(a_m)}.
\end{equation} This gives \cref{eq:gain}. The metric combines correlation and relative amplitude. It is not an energy fraction and can therefore remain appreciable when the translation template explains little of the total unconditioned variance.

The memory ratios in \cref{eq:passage} use only body-scale freestream normalization. At $\KnD=0.01$, $\deltaten/D=0.0615$ and $\scenter/D=0.231$, giving $\tau_p^*/t_\delta^*=10.61$ and $\tau_p^*/t_s^*=2.83$. At $\KnD=0.025$, the corresponding ratios are $6.46$ and $3.09$. These values are descriptive geometric comparisons, not residence times, universal constants or fitted scaling laws.

\section{Separation from the companion mean-flow analysis}
The companion manuscript \citep{RoohiShoja2026Mean} uses converged mean fields across Mach- and Knudsen-number sweeps. Its principal observables are mean front location, mean thickness, variable-specific static relaxation lengths, profile registration and proper orthogonal decomposition (POD) across operating conditions. The present paper uses consecutive outputs at fixed operating conditions. Its principal observables are temporal marker covariance, autoregressive memory, angular eigenfunctions, full-field equivalent displacement amplitudes, sliding-window persistence and injection-based detectability. No temporal covariance, memory, coherent-gain or multi-moment synchronization result in the present paper is reported in the companion work.

\FloatBarrier

\bibliographystyle{jfm}
\bibliography{references}

@book{Bird1994,
  author    = {Bird, Graeme A.},
  title     = {Molecular Gas Dynamics and the Direct Simulation of Gas Flows},
  publisher = {Clarendon Press},
  address   = {Oxford},
  year      = {1994},
  doi       = {10.1093/oso/9780198561958.001.0001}
}

@article{Stefanov2000,
  author  = {Stefanov, Stefan K. and Boyd, Iain D. and Cai, Chunpei},
  title   = {Monte Carlo analysis of macroscopic fluctuations in a rarefied hypersonic flow around a cylinder},
  journal = {Physics of Fluids},
  volume  = {12},
  number  = {5},
  pages   = {1226--1239},
  year    = {2000},
  doi     = {10.1063/1.870372}
}

@article{Lofthouse2007,
  author  = {Lofthouse, Andrew J. and Boyd, Iain D. and Wright, Michael J.},
  title   = {Effects of continuum breakdown on hypersonic aerothermodynamics},
  journal = {Physics of Fluids},
  volume  = {19},
  number  = {2},
  pages   = {027105},
  year    = {2007},
  doi     = {10.1063/1.2710289}
}

@article{Akhlaghi2017,
  author  = {Akhlaghi, Hassan and Daliri, Abbas and Soltani, Mohammad Reza},
  title   = {Shock-wave-detection technique for high-speed rarefied-gas flows},
  journal = {AIAA Journal},
  volume  = {55},
  number  = {11},
  pages   = {3747--3756},
  year    = {2017},
  doi     = {10.2514/1.J055705}
}

@article{Akhlaghi2021,
  author  = {Akhlaghi, Hassan and Roohi, Ehsan and Daliri, Abbas and Soltani, Mohammad-Reza},
  title   = {Shock polar investigation in supersonic rarefied gas flows over a circular cylinder},
  journal = {Physics of Fluids},
  volume  = {33},
  number  = {5},
  pages   = {052006},
  year    = {2021},
  doi     = {10.1063/5.0049516}
}

@article{Plotkin1975,
  author  = {Plotkin, Kenneth J.},
  title   = {Shock wave oscillation driven by turbulent boundary-layer fluctuations},
  journal = {AIAA Journal},
  volume  = {13},
  number  = {8},
  pages   = {1036--1040},
  year    = {1975},
  doi     = {10.2514/3.60501}
}

@article{PoggieSmits2001,
  author  = {Poggie, Jonathan and Smits, Alexander J.},
  title   = {Shock unsteadiness in a reattaching shear layer},
  journal = {Journal of Fluid Mechanics},
  volume  = {429},
  pages   = {155--185},
  year    = {2001},
  doi     = {10.1017/S002211200000269X}
}

@article{PoggieSmits2005,
  author  = {Poggie, Jonathan and Smits, Alexander J.},
  title   = {Experimental evidence for Plotkin model of shock unsteadiness in separated flow},
  journal = {Physics of Fluids},
  volume  = {17},
  number  = {1},
  pages   = {018107},
  year    = {2005},
  doi     = {10.1063/1.1847412}
}

@article{Dussauge2006,
  author  = {Dussauge, Jean-Paul and Dupont, Pierre and Debi{\`e}ve, Jean-Fran{\c{c}}ois},
  title   = {Unsteadiness in shock wave boundary layer interactions with separation},
  journal = {Aerospace Science and Technology},
  volume  = {10},
  number  = {2},
  pages   = {85--91},
  year    = {2006},
  doi     = {10.1016/j.ast.2005.09.001}
}

@article{Dupont2006,
  author  = {Dupont, Pierre and Haddad, Chadi and Debi{\`e}ve, Jean-Fran{\c{c}}ois},
  title   = {Space and time organization in a shock-induced separated boundary layer},
  journal = {Journal of Fluid Mechanics},
  volume  = {559},
  pages   = {255--277},
  year    = {2006},
  doi     = {10.1017/S0022112006000267}
}

@article{Piponniau2009,
  author  = {Piponniau, S. and Dussauge, J.-P. and Debi{\`e}ve, J.-F. and Dupont, P.},
  title   = {A simple model for low-frequency unsteadiness in shock-induced separation},
  journal = {Journal of Fluid Mechanics},
  volume  = {629},
  pages   = {87--108},
  year    = {2009},
  doi     = {10.1017/S0022112009006417}
}

@article{TouberSandham2011,
  author  = {Touber, Emile and Sandham, Neil D.},
  title   = {Low-order stochastic modelling of low-frequency motions in reflected shock-wave/boundary-layer interactions},
  journal = {Journal of Fluid Mechanics},
  volume  = {671},
  pages   = {417--465},
  year    = {2011},
  doi     = {10.1017/S0022112010005811}
}

@article{PriebeMartin2012,
  author  = {Priebe, Stephan and Martin, M. Pino},
  title   = {Low-frequency unsteadiness in shock wave--turbulent boundary layer interaction},
  journal = {Journal of Fluid Mechanics},
  volume  = {699},
  pages   = {1--49},
  year    = {2012},
  doi     = {10.1017/jfm.2011.560}
}

@article{ClemensNarayanaswamy2014,
  author  = {Clemens, Noel T. and Narayanaswamy, Venkateswaran},
  title   = {Low-frequency unsteadiness of shock wave/turbulent boundary layer interactions},
  journal = {Annual Review of Fluid Mechanics},
  volume  = {46},
  pages   = {469--492},
  year    = {2014},
  doi     = {10.1146/annurev-fluid-010313-141346}
}

@article{OrnsteinUhlenbeck1930,
  author  = {Uhlenbeck, George E. and Ornstein, Leonard S.},
  title   = {On the theory of the Brownian motion},
  journal = {Physical Review},
  volume  = {36},
  number  = {5},
  pages   = {823--841},
  year    = {1930},
  doi     = {10.1103/PhysRev.36.823}
}

@article{Hadjiconstantinou2003,
  author  = {Hadjiconstantinou, Nicolas G. and Garcia, Alejandro L. and Bazant, Martin Z. and He, Gang},
  title   = {Statistical error in particle simulations of hydrodynamic phenomena},
  journal = {Journal of Computational Physics},
  volume  = {187},
  number  = {1},
  pages   = {274--297},
  year    = {2003},
  doi     = {10.1016/S0021-9991(03)00099-8}
}

@article{BellGarciaWilliams2007,
  author  = {Bell, John B. and Garcia, Alejandro L. and Williams, Sarah A.},
  title   = {Numerical methods for the stochastic Landau--Lifshitz Navier--Stokes equations},
  journal = {Physical Review E},
  volume  = {76},
  pages   = {016708},
  year    = {2007},
  doi     = {10.1103/PhysRevE.76.016708}
}

@article{WilliamsBellGarcia2008,
  author  = {Williams, Sarah A. and Bell, John B. and Garcia, Alejandro L.},
  title   = {Algorithm refinement for fluctuating hydrodynamics},
  journal = {Multiscale Modeling and Simulation},
  volume  = {6},
  number  = {4},
  pages   = {1256--1280},
  year    = {2008},
  doi     = {10.1137/070696180}
}

@article{Gallis2016,
  author  = {Gallis, M. A. and Koehler, T. P. and Torczynski, J. R. and Plimpton, S. J.},
  title   = {Direct simulation Monte Carlo investigation of the Rayleigh--Taylor instability},
  journal = {Physical Review Fluids},
  volume  = {1},
  pages   = {043403},
  year    = {2016},
  doi     = {10.1103/PhysRevFluids.1.043403}
}

@article{Gallis2017,
  author  = {Gallis, M. A. and Bitter, N. P. and Koehler, T. P. and Torczynski, J. R. and Plimpton, S. J. and Papadakis, G.},
  title   = {Molecular-level simulations of turbulence and its decay},
  journal = {Physical Review Letters},
  volume  = {118},
  pages   = {064501},
  year    = {2017},
  doi     = {10.1103/PhysRevLett.118.064501}
}

@article{Gallis2021,
  author  = {Gallis, M. A. and Torczynski, J. R. and Krygier, M. C. and Bitter, N. P. and Plimpton, S. J.},
  title   = {Turbulence at the edge of continuum},
  journal = {Physical Review Fluids},
  volume  = {6},
  pages   = {013401},
  year    = {2021},
  doi     = {10.1103/PhysRevFluids.6.013401}
}

@article{Bell2022,
  author  = {Bell, John B. and Nonaka, Andrew and Garcia, Alejandro L. and Eyink, Gregory},
  title   = {Thermal fluctuations in the dissipation range of homogeneous isotropic turbulence},
  journal = {Journal of Fluid Mechanics},
  volume  = {939},
  pages   = {A12},
  year    = {2022},
  doi     = {10.1017/jfm.2022.188}
}

@article{Kunsch1989,
  author  = {K{\"u}nsch, Hans R.},
  title   = {The jackknife and the bootstrap for general stationary observations},
  journal = {The Annals of Statistics},
  volume  = {17},
  number  = {3},
  pages   = {1217--1241},
  year    = {1989},
  doi     = {10.1214/aos/1176347265}
}

@article{PolitisRomano1994,
  author  = {Politis, Dimitris N. and Romano, Joseph P.},
  title   = {The stationary bootstrap},
  journal = {Journal of the American Statistical Association},
  volume  = {89},
  number  = {428},
  pages   = {1303--1313},
  year    = {1994},
  doi     = {10.1080/01621459.1994.10476870}
}

@article{Higham2002,
  author  = {Higham, Nicholas J.},
  title   = {Computing the nearest correlation matrix---a problem from finance},
  journal = {IMA Journal of Numerical Analysis},
  volume  = {22},
  number  = {3},
  pages   = {329--343},
  year    = {2002},
  doi     = {10.1093/imanum/22.3.329}
}

@article{Welch1967,
  author  = {Welch, Peter D.},
  title   = {The use of fast Fourier transform for the estimation of power spectra: a method based on time averaging over short, modified periodograms},
  journal = {IEEE Transactions on Audio and Electroacoustics},
  volume  = {15},
  number  = {2},
  pages   = {70--73},
  year    = {1967},
  doi     = {10.1109/TAU.1967.1161901}
}

@article{Lumley1967,
  author  = {Lumley, John L.},
  title   = {The structure of inhomogeneous turbulent flows},
  journal = {Atmospheric Turbulence and Radio Wave Propagation},
  pages   = {166--178},
  year    = {1967}
}

@article{Sirovich1987,
  author  = {Sirovich, Lawrence},
  title   = {Turbulence and the dynamics of coherent structures. Part I: coherent structures},
  journal = {Quarterly of Applied Mathematics},
  volume  = {45},
  number  = {3},
  pages   = {561--571},
  year    = {1987},
  doi     = {10.1090/qam/910462}
}

@article{Berkooz1993,
  author  = {Berkooz, Gal and Holmes, Philip and Lumley, John L.},
  title   = {The proper orthogonal decomposition in the analysis of turbulent flows},
  journal = {Annual Review of Fluid Mechanics},
  volume  = {25},
  pages   = {539--575},
  year    = {1993},
  doi     = {10.1146/annurev.fl.25.010193.002543}
}

@article{Taira2017,
  author  = {Taira, Kunihiko and Brunton, Steven L. and Dawson, Scott T. M. and Rowley, Clarence W. and Colonius, Tim and McKeon, Beverley J. and Schmidt, Oliver T. and Gordeyev, Stanislav and Theofilis, Vassilios and Ukeiley, Lawrence S.},
  title   = {Modal analysis of fluid flows: an overview},
  journal = {AIAA Journal},
  volume  = {55},
  number  = {12},
  pages   = {4013--4041},
  year    = {2017},
  doi     = {10.2514/1.J056060}
}

@article{Schmid2010,
  author  = {Schmid, Peter J.},
  title   = {Dynamic mode decomposition of numerical and experimental data},
  journal = {Journal of Fluid Mechanics},
  volume  = {656},
  pages   = {5--28},
  year    = {2010},
  doi     = {10.1017/S0022112010001217}
}

@article{Towne2018,
  author  = {Towne, Aaron and Schmidt, Oliver T. and Colonius, Tim},
  title   = {Spectral proper orthogonal decomposition and its relationship to dynamic mode decomposition and resolvent analysis},
  journal = {Journal of Fluid Mechanics},
  volume  = {847},
  pages   = {821--867},
  year    = {2018},
  doi     = {10.1017/jfm.2018.283}
}

@article{SchmidtColonius2020,
  author  = {Schmidt, Oliver T. and Colonius, Tim},
  title   = {Guide to spectral proper orthogonal decomposition},
  journal = {AIAA Journal},
  volume  = {58},
  number  = {3},
  pages   = {1023--1033},
  year    = {2020},
  doi     = {10.2514/1.J058809}
}

@article{RowleyMarsden2000,
  author  = {Rowley, Clarence W. and Marsden, Jerrold E.},
  title   = {Reconstruction equations and the Karhunen--Lo{\`e}ve expansion for systems with symmetry},
  journal = {Physica D},
  volume  = {142},
  pages   = {1--19},
  year    = {2000},
  doi     = {10.1016/S0167-2789(00)00042-7}
}

@article{Reiss2018,
  author  = {Reiss, Julius and Schulze, Philipp and Sesterhenn, J{\"o}rn and Mehrmann, Volker},
  title   = {The shifted proper orthogonal decomposition: a mode decomposition for multiple transport phenomena},
  journal = {SIAM Journal on Scientific Computing},
  volume  = {40},
  number  = {3},
  pages   = {A1322--A1344},
  year    = {2018},
  doi     = {10.1137/17M1140571}
}

@article{Taddei2020,
  author  = {Taddei, Tommaso},
  title   = {A registration method for model order reduction: data compression and geometry reduction},
  journal = {SIAM Journal on Scientific Computing},
  volume  = {42},
  number  = {2},
  pages   = {A997--A1027},
  year    = {2020},
  doi     = {10.1137/19M1271270}
}

@article{Sawant2021,
  author  = {Sawant, Saurabh S. and Levin, Deborah A. and Theofilis, Vassilis},
  title   = {A kinetic approach to studying low-frequency molecular fluctuations in a one-dimensional shock},
  journal = {Physics of Fluids},
  volume  = {33},
  pages   = {104106},
  year    = {2021},
  doi     = {10.1063/5.0065971}
}

@inproceedings{Senkardesler2026,
  author    = {Senkardesler, Mert and Karpuzcu, Irmak T. and Levin, Deborah A.},
  title     = {{DSMC} study on unsteadiness of supersonic flow over a cylinder},
  booktitle = {AIAA SCITECH 2026 Forum},
  publisher = {American Institute of Aeronautics and Astronautics},
  year      = {2026},
  note      = {AIAA Paper 2026-2147},
  doi       = {10.2514/6.2026-2147}
}

@article{Sawant2022,
  author  = {Sawant, Saurabh S. and Theofilis, Vassilis and Levin, Deborah A.},
  title   = {Analytical prediction of low-frequency fluctuations inside a one-dimensional shock},
  journal = {Physics of Fluids},
  volume  = {34},
  pages   = {066103},
  year    = {2022},
  doi     = {10.1063/5.0091320}
}

@article{Klothakis2022,
  author  = {Klothakis, Angelos and Quintanilha, Helio and Sawant, Saurabh S. and Protopapadakis, Eftychios and Theofilis, Vassilis and Levin, Deborah A.},
  title   = {Linear stability analysis of hypersonic boundary layers computed by a kinetic approach: a semi-infinite flat plate at Mach 4.5 and 9},
  journal = {Physics of Fluids},
  volume  = {34},
  pages   = {034105},
  year    = {2022},
  doi     = {10.1063/5.0065150}
}

@article{Karpuzcu2025,
  author  = {Karpuzcu, Irmak T. and Senkardesler, Mert and Levin, Deborah A.},
  title   = {On flow unsteadiness in strongly separated high-speed ramp flows using kinetic and data-driven methods},
  journal = {Physics of Fluids},
  volume  = {37},
  pages   = {096136},
  year    = {2025},
  doi     = {10.1063/5.0288620}
}

@misc{Karpuzcu2026,
  author        = {Karpuzcu, Irmak T. and Levin, Deborah A. and Theofilis, Vassilis},
  title         = {A kinetic linear stability theory framework for high speed flows},
  year          = {2026},
  eprint        = {2607.27440},
  archivePrefix = {arXiv},
  primaryClass  = {physics.flu-dyn},
  doi           = {10.48550/arXiv.2607.27440}
}

@article{Zou2023,
  author  = {Zou, S. and Bi, L. and Zhong, C. and Yuan, X. and Tang, Z.},
  title   = {A novel linear stability analysis method for plane Couette flow considering rarefaction effects},
  journal = {Journal of Fluid Mechanics},
  volume  = {963},
  pages   = {A33},
  year    = {2023},
  doi     = {10.1017/jfm.2023.289}
}

@article{MillikanWhite1963,
  author  = {Millikan, Roger C. and White, Donald R.},
  title   = {Systematics of vibrational relaxation},
  journal = {The Journal of Chemical Physics},
  volume  = {39},
  number  = {12},
  pages   = {3209--3213},
  year    = {1963},
  doi     = {10.1063/1.1734182}
}

@article{Park1988,
  author  = {Park, Chul},
  title   = {Assessment of a two-temperature kinetic model for dissociating and weakly ionizing nitrogen},
  journal = {Journal of Thermophysics and Heat Transfer},
  volume  = {2},
  number  = {1},
  pages   = {8--16},
  year    = {1988},
  doi     = {10.2514/3.55}
}

@article{Bertolotti1998,
  author  = {Bertolotti, Fabio P.},
  title   = {The influence of rotational and vibrational energy relaxation on boundary-layer stability},
  journal = {Journal of Fluid Mechanics},
  volume  = {372},
  pages   = {93--118},
  year    = {1998},
  doi     = {10.1017/S0022112098002353}
}

@misc{RoohiShoja2026Mean,
  author        = {Roohi, Ehsan and Shoja-Sani, Ahmad},
  title         = {Rarefaction-induced inflation and similarity breakdown of hypersonic bow shocks over a circular cylinder},
  year          = {2026},
  eprint        = {2605.17099},
  archivePrefix = {arXiv},
  primaryClass  = {physics.flu-dyn},
  doi           = {10.48550/arXiv.2605.17099},
  note          = {arXiv preprint arXiv:2605.17099}
}

@article{RoohiShojaAzghadi2026PoF,
  author  = {Roohi, Ehsan and Shoja-Sani, Ahmad and Ebrahimzadeh Azghadi, Fahimeh},
  title   = {Neural networks for rarefied gas dynamics: Relaxation problem, polyatomic shock waves, and hypersonic cylinder flow},
  journal = {Physics of Fluids},
  volume  = {38},
  number  = {5},
  pages   = {057108},
  year    = {2026},
  doi     = {10.1063/5.0334590}
}
\end{document}